\RequirePackage{lineno}  

\documentclass[prd,aps,superscriptaddress,showpacs,amsmath,amssymb]{revtex4}

\usepackage{graphicx}
\usepackage{dcolumn}
\usepackage{bm}
\usepackage[dvips]{epsfig}

\begin{document}

\newcommand{\xkshort}     {$K_S^0$} 
\newcommand{\xkstar}       {$K^{\star\pm}$} 
\newcommand{\xpi}              {$\pi^{\pm}$}
\newcommand{\xphi}           {$\phi$}
\newcommand{\xlambda}   {$\Lambda^0$} 
\newcommand{\xlambdab}  {$\overline{\Lambda}^0$}
\newcommand{\xet}             {$E_T$} 
\newcommand{\xptmin}      {$p_T^{min}} 
\newcommand{\x}              {$\times$} 

\title{Production of $K_S^0$, $K^{\star\pm} (892)$ and $\phi^0 (1020)$  in minimum bias events  and $K_S^0$ and $\Lambda^{0}$ in jets in $p\overline{p}$ collisions at $\sqrt{s}=1.96$ TeV}
\date{\today}
\affiliation{Institute of Physics, Academia Sinica, Taipei, Taiwan 11529, Republic of China}
\affiliation{Argonne National Laboratory, Argonne, Illinois 60439, USA}
\affiliation{University of Athens, 157 71 Athens, Greece}
\affiliation{Institut de Fisica d'Altes Energies, ICREA, Universitat Autonoma de Barcelona, E-08193, Bellaterra (Barcelona), Spain}
\affiliation{Baylor University, Waco, Texas 76798, USA}
\affiliation{Istituto Nazionale di Fisica Nucleare Bologna, $^{ee}$University of Bologna, I-40127 Bologna, Italy}
\affiliation{University of California, Davis, Davis, California 95616, USA}
\affiliation{University of California, Los Angeles, Los Angeles, California 90024, USA}
\affiliation{Instituto de Fisica de Cantabria, CSIC-University of Cantabria, 39005 Santander, Spain}
\affiliation{Carnegie Mellon University, Pittsburgh, Pennsylvania 15213, USA}
\affiliation{Enrico Fermi Institute, University of Chicago, Chicago, Illinois 60637, USA}
\affiliation{Comenius University, 842 48 Bratislava, Slovakia; Institute of Experimental Physics, 040 01 Kosice, Slovakia}
\affiliation{Joint Institute for Nuclear Research, RU-141980 Dubna, Russia}
\affiliation{Duke University, Durham, North Carolina 27708, USA}
\affiliation{Fermi National Accelerator Laboratory, Batavia, Illinois 60510, USA}
\affiliation{University of Florida, Gainesville, Florida 32611, USA}
\affiliation{Laboratori Nazionali di Frascati, Istituto Nazionale di Fisica Nucleare, I-00044 Frascati, Italy}
\affiliation{University of Geneva, CH-1211 Geneva 4, Switzerland}
\affiliation{Glasgow University, Glasgow G12 8QQ, United Kingdom}
\affiliation{Harvard University, Cambridge, Massachusetts 02138, USA}
\affiliation{Division of High Energy Physics, Department of Physics, University of Helsinki and Helsinki Institute of Physics, FIN-00014, Helsinki, Finland}
\affiliation{University of Illinois, Urbana, Illinois 61801, USA}
\affiliation{The Johns Hopkins University, Baltimore, Maryland 21218, USA}
\affiliation{Institut f\"{u}r Experimentelle Kernphysik, Karlsruhe Institute of Technology, D-76131 Karlsruhe, Germany}
\affiliation{Center for High Energy Physics: Kyungpook National University, Daegu 702-701, Korea; Seoul National University, Seoul 151-742, Korea; Sungkyunkwan University, Suwon 440-746, Korea; Korea Institute of Science and Technology Information, Daejeon 305-806, Korea; Chonnam National University, Gwangju 500-757, Korea; Chonbuk National University, Jeonju 561-756, Korea}
\affiliation{Ernest Orlando Lawrence Berkeley National Laboratory, Berkeley, California 94720, USA}
\affiliation{University of Liverpool, Liverpool L69 7ZE, United Kingdom}
\affiliation{University College London, London WC1E 6BT, United Kingdom}
\affiliation{Centro de Investigaciones Energeticas Medioambientales y Tecnologicas, E-28040 Madrid, Spain}
\affiliation{Massachusetts Institute of Technology, Cambridge, Massachusetts 02139, USA}
\affiliation{Institute of Particle Physics: McGill University, Montr\'{e}al, Qu\'{e}bec, Canada H3A~2T8; Simon Fraser University, Burnaby, British Columbia, Canada V5A~1S6; University of Toronto, Toronto, Ontario, Canada M5S~1A7; and TRIUMF, Vancouver, British Columbia, Canada V6T~2A3}
\affiliation{University of Michigan, Ann Arbor, Michigan 48109, USA}
\affiliation{Michigan State University, East Lansing, Michigan 48824, USA}
\affiliation{Institution for Theoretical and Experimental Physics, ITEP, Moscow 117259, Russia}
\affiliation{University of New Mexico, Albuquerque, New Mexico 87131, USA}
\affiliation{The Ohio State University, Columbus, Ohio 43210, USA}
\affiliation{Okayama University, Okayama 700-8530, Japan}
\affiliation{Osaka City University, Osaka 588, Japan}
\affiliation{University of Oxford, Oxford OX1 3RH, United Kingdom}
\affiliation{Istituto Nazionale di Fisica Nucleare, Sezione di Padova-Trento, $^{ff}$University of Padova, I-35131 Padova, Italy}
\affiliation{University of Pennsylvania, Philadelphia, Pennsylvania 19104, USA}
\affiliation{Istituto Nazionale di Fisica Nucleare Pisa, $^{gg}$University of Pisa, $^{hh}$University of Siena and $^{ii}$Scuola Normale Superiore, I-56127 Pisa, Italy}
\affiliation{University of Pittsburgh, Pittsburgh, Pennsylvania 15260, USA}
\affiliation{Purdue University, West Lafayette, Indiana 47907, USA}
\affiliation{University of Rochester, Rochester, New York 14627, USA}
\affiliation{The Rockefeller University, New York, New York 10065, USA}
\affiliation{Istituto Nazionale di Fisica Nucleare, Sezione di Roma 1, $^{jj}$Sapienza Universit\`{a} di Roma, I-00185 Roma, Italy}
\affiliation{Rutgers University, Piscataway, New Jersey 08855, USA}
\affiliation{Texas A\&M University, College Station, Texas 77843, USA}
\affiliation{Istituto Nazionale di Fisica Nucleare Trieste/Udine, I-34100 Trieste, $^{kk}$University of Udine, I-33100 Udine, Italy}
\affiliation{University of Tsukuba, Tsukuba, Ibaraki 305, Japan}
\affiliation{Tufts University, Medford, Massachusetts 02155, USA}
\affiliation{University of Virginia, Charlottesville, Virginia 22906, USA}
\affiliation{Waseda University, Tokyo 169, Japan}
\affiliation{Wayne State University, Detroit, Michigan 48201, USA}
\affiliation{University of Wisconsin, Madison, Wisconsin 53706, USA}
\affiliation{Yale University, New Haven, Connecticut 06520, USA}

\author{T.~Aaltonen}
\affiliation{Division of High Energy Physics, Department of Physics, University of Helsinki and Helsinki Institute of Physics, FIN-00014, Helsinki, Finland}
\author{M.~Albrow}
\affiliation{Fermi National Accelerator Laboratory, Batavia, Illinois 60510, USA}
\author{B.~\'{A}lvarez~Gonz\'{a}lez$^z$}
\affiliation{Instituto de Fisica de Cantabria, CSIC-University of Cantabria, 39005 Santander, Spain}
\author{S.~Amerio}
\affiliation{Istituto Nazionale di Fisica Nucleare, Sezione di Padova-Trento, $^{ff}$University of Padova, I-35131 Padova, Italy}
\author{D.~Amidei}
\affiliation{University of Michigan, Ann Arbor, Michigan 48109, USA}
\author{A.~Anastassov$^x$}
\affiliation{Fermi National Accelerator Laboratory, Batavia, Illinois 60510, USA}
\author{A.~Annovi}
\affiliation{Laboratori Nazionali di Frascati, Istituto Nazionale di Fisica Nucleare, I-00044 Frascati, Italy}
\author{J.~Antos}
\affiliation{Comenius University, 842 48 Bratislava, Slovakia; Institute of Experimental Physics, 040 01 Kosice, Slovakia}
\author{G.~Apollinari}
\affiliation{Fermi National Accelerator Laboratory, Batavia, Illinois 60510, USA}
\author{J.A.~Appel}
\affiliation{Fermi National Accelerator Laboratory, Batavia, Illinois 60510, USA}
% ...

\collaboration{CDF Collaboration\footnote{With visitors from
$^a$Istituto Nazionale di Fisica Nucleare, Sezione di Cagliari, 09042 Monserrato (Cagliari), Italy,
$^b$University of CA Irvine, Irvine, CA 92697, USA,
$^c$University of CA Santa Barbara, Santa Barbara, CA 93106, USA,
$^d$University of CA Santa Cruz, Santa Cruz, CA 95064, USA,
$^e$Institute of Physics, Academy of Sciences of the Czech Republic, Czech Republic,
$^f$CERN, CH-1211 Geneva, Switzerland,
$^g$Cornell University, Ithaca, NY 14853, USA,
$^h$University of Cyprus, Nicosia CY-1678, Cyprus,
$^i$Office of Science, U.S. Department of Energy, Washington, DC 20585, USA,
$^j$University College Dublin, Dublin 4, Ireland,
$^k$ETH, 8092 Zurich, Switzerland,
$^l$University of Fukui, Fukui City, Fukui Prefecture, Japan 910-0017,
$^m$Universidad Iberoamericana, Mexico D.F., Mexico,
$^n$University of Iowa, Iowa City, IA 52242, USA,
$^o$Kinki University, Higashi-Osaka City, Japan 577-8502,
$^p$Kansas State University, Manhattan, KS 66506, USA,
$^q$Korea University, Seoul, 136-713, Korea,
$^r$University of Manchester, Manchester M13 9PL, United Kingdom,
$^s$Queen Mary, University of London, London, E1 4NS, United Kingdom,
$^t$University of Melbourne, Victoria 3010, Australia,
$^u$Muons, Inc., Batavia, IL 60510, USA,
$^v$Nagasaki Institute of Applied Science, Nagasaki, Japan,
$^w$National Research Nuclear University, Moscow, Russia,
$^x$Northwestern University, Evanston, IL 60208, USA,
$^y$University of Notre Dame, Notre Dame, IN 46556, USA,
$^z$Universidad de Oviedo, E-33007 Oviedo, Spain,
$^{aa}$CNRS-IN2P3, Paris, F-75205 France,
$^{bb}$Texas Tech University, Lubbock, TX 79609, USA,
$^{cc}$Universidad Tecnica Federico Santa Maria, 110v Valparaiso, Chile,
$^{dd}$Yarmouk University, Irbid 211-63, Jordan,
}}
\noaffiliation



\begin{abstract}
We report measurements of the inclusive transverse momentum ($p_T$) distribution of centrally produced $K_S^0$, $K^{\star\pm} (892)$, and $\phi^0 (1020)$  mesons up to $p_T$ = 10 GeV/$c$  in minimum-bias events, and  $K_S^0$ and $\Lambda^{0}$ particles up to   $p_T$ = 20 GeV/c in jets with transverse energy between 25 GeV and 160 GeV in $p \bar p$ collisions. The data were taken with the CDF II detector at the Fermilab Tevatron at $\sqrt{s}$ = 1.96 TeV. We find that as $p_T$ increases, the $p_T$ slopes of the three mesons ($K_S^0$, $K^{\star\pm}$, and $\phi$) are similar, and the ratio of $\Lambda^{0}$ to $K_S^0$ as a function of $p_T$ in minimum-bias events becomes similar to the fairly constant ratio in jets at $p_T\sim$ 5 GeV/c. This suggests that the particles with $p_T\gtrsim$ 5 GeV/c in minimum-bias events are from ``soft" jets, and that the $p_T$ slope of particles in jets is insensitive to light quark flavor ($u$, $d$, or $s$) and to the number of valence quarks. We also find that for $p_T \lesssim$ 4 GeV relatively more \xlambda~baryons are produced  in minimum-bias events than in jets. 
\end{abstract}

\maketitle
\section{INTRODUCTION}
The study of particles with low $p_T$  (transverse momentum with respect to the beam direction) from 
hadron-hadron interactions is as old as high energy physics itself. Nevertheless, attempts  to understand the physics of 
particle production have had limited success. 
As the center-of-mass energy increases, the number of produced particles increases and events get 
more complex.  Although the discovery of high transverse energy, $E_T$, jets in hadron collisions at the 
CERN ISR~\cite{ref:isrjet} and $S{\sc{p\overline{p}}}S$ Collider~\cite{ref:ua1jet} 
supported the theory of strongly interacting quarks and gluons (QCD), low $p_T$ hadron production
is still not well understood despite additional data from $p\overline{p}$ and $pp$ colliders 
including RHIC~\cite{ref:hhcollision} because the strong coupling is large, and perturbative QCD calculations do not apply. 
Phenomenological models, such as \textsc{pythia}~\cite{ref:pythia}, have been developed and tuned to data. 
New data, such as that presented here on strange particle production, can further refine the models.

Hadron-hadron collisions are classified into two types, elastic and inelastic collisions. Inelastic hadron-hadron collisions
are generally further classified as diffractive  and non-diffractive. The diffractive events have a large rapidity 
gap ($>$~3) with no hadrons. The distinction is not absolute, and experiments (and theorists) should make their definitions explicit.

Inelastic collisions can have a hard parton-parton interaction resulting in high $E_T$ jets, and we select events with 
jets with $E_T$ from 25 to 160 GeV and measure the production of hadrons with strange quarks in the jets. In this paper, 
we present the invariant differential cross section, $E d^3\sigma/dp^3$, of $K_S^0$, $K^{\star\pm}$, 
and $\phi$ particles up to $p_T$ = 10 GeV/c in typical non-diffractive events, and the $p_T$ distributions 
of \xkshort~and \xlambda~in jets up to $p_T$ = 20 GeV/c and 
jet $E_T$ = 160 GeV. This is the first time that the $p_T$ distributions of identified particles in high-$E_T$ jets from 
hadron-hadron collisions have been measured. These spectra extend down to $p_T \sim$ 1 GeV/c where perturbative 
calculations cannot be used. The various phenomenological approaches in this region (some inspired by QCD)
 benefit from such data.

One goal of this analysis is to compare particle production from minimum-bias (MB) and jet events to see if there is a transition
at some $p_T$, above which the particles from jet fragmentation tend to dominate. Another goal is to test the fragmentation 
process of quarks and gluons to jets in the {\sc pythia} event generator tuned to  $e^+e^-$~\cite{ref:eedata} and 
$e^-p$~\cite{ref:epdata} data. Because the particles are identified, the comparison can be more sensitive to 
details, e.g., $s$-quark creation.  A third goal is to provide information on particles produced with $p_T$ less 
than $\sim 3$ GeV/$c$ in MB events. Apart from their intrinsic
interest, such data are useful in searches for quark-gluon plasma signatures in heavy-ion collisions. 
 
\section{EVENT AND JET  SELECTION}
The data in this analysis are from the CDF II detector at the Tevatron Collider operating at a center-of-mass energy 
$\sqrt{s}$ = 1.96 TeV. The CDF II detector was described in detail elsewhere~\cite{ref:cdfdet}.  The components most 
relevant to this analysis are the  tracking system and the calorimeters.  The tracking system 
was in a uniform axial magnetic field of $1.4$ T.  The inner tracker had seven to eight 
layers of  silicon microstrip detectors ranging in radius from 1.5 to 28.0 cm~\cite{ref:cdfsvx} in the 
pseudorapidity region $|\eta|<2$~\cite{ref:cdfxyz}. Outside this was 
the Central Outer Tracker (COT) a cylindrical drift chamber with 96 sense-wire layers  
grouped in eight superlayers of axial and stereo wires~\cite{ref:cdfcot}. Its active volume 
covered 40 to 140 cm in radius and $|z|<155$ cm, where $z$ is the coordinate along the beam direction
centered in the middle of the detector.

Surrounding the tracking system were the pointing-tower-geometry electromagnetic (EM) and hadronic calorimeters ~\cite{ref:cdfcal1}, 
divided into central ($|\eta|<1.1$) and plug (1.1 $<|\eta|<$ 3.6) regions. The calorimeters were 
made of lead (EM) and iron (hadronic) absorbers sandwiched between plastic scintillators that provided 
measurements of shower energies. At a depth approximately corresponding to the maximum development of the 
typical electromagnetic shower, the EM calorimeters contained
proportional chambers~\cite{ref:cdfcal2} to measure shower positions and profiles.

MB events were collected with a trigger selecting beam-bunch
crossings with at least one inelastic $p\bar{p}$ interaction. We required a time coincidence between signals  in 
both  forward and backward gas Cherenkov counters~\cite{ref:cdfclc} covering the regions $3.7 < |\eta| < 4.7$.  
In these events we study $K_S^0$, $K^{\star\pm}$, and $\phi$ production in the central region, $|\eta| < $ 1.0. 

The high-$E_T$ jet events were collected with four jet transverse-energy trigger thresholds: 20, 50, 
70, and 100 GeV, and the lower $E_T$ threshold events were randomly accepted at a fixed fraction in order
to reduce the trigger rate. Jets are constructed using a fixed-cone algorithm with radius 
$\Delta R = \sqrt{(\Delta\eta)^2 + (\Delta\phi)^2}$ = 0.4, and their energies are corrected for detector 
effects ~\cite{ref:cdfcal3}. Jets with $|\eta| <$ 1.0 are used and these jets are divided into five $E_T$ ranges: 
25 -- 40 GeV, 40 -- 60 GeV, 60 -- 80 GeV, 80 -- 120 GeV and 120 -- 160 GeV. 
We study the production properties of \xkshort~and \xlambda~ for each range.   

We require a reconstructed event vertex in the fiducial region $|z_{\rm VTX}| \le 60$ cm.
Tracks are required to have a high track-fit quality, with $\chi^2$ per degree-of-freedom 
($\chi^2/$dof)~$ \le 2.5$, with more than five hits in at least two axial and two stereo COT track segments
reconstructed in superlayers. It is further required that tracks have $|\eta|<1$ 
and $p_T > p_T^{\rm min}$, where $p_T^{\rm min}$ = 0.325 GeV/c and 0.5 GeV/c for MB events and jet events 
respectively.
 
\section{\xkshort~and \xlambda~reconstruction}
The \xkshort~and \xlambda~reconstruction procedures are similar. Since the \xlambda~
reconstruction is well described in a previous publication ~\cite{ref:cdfhyp}, 
a summary for \xkshort~reconstruction is presented here.  We search for \xkshort~to $\pi^+\pi^-$ 
decays using tracks with opposite charge and  $p_T > p_T^{\rm min}$ that satisfy the $\chi^2/$dof 
and COT segment requirements.  

For each track pair we calculate the position of their intersection in the transverse ($r-\phi$) plane. 
Once this intersection point, referred to as the secondary vertex, is found, the $z$-coordinate 
of each track ($z_{1}$ and $z_{2}$) is calculated at that point. If the distance $|z_1 - z_2|$ is 
less than 1.5 cm, the tracks are considered to originate from a \xkshort~candidate decay. The pair 
is traced  back to the primary event vertex and we require $\delta z_0$ to be less than $2$ cm,  and $d_0$ to 
be less than $0.25$ cm. The quantities $\delta z_0$ and $d_0$ are the distances between the event vertex and the 
track position at the point of closest approach to the event vertex in the  $z$-axis and in 
the $r-\phi$ plane respectively. To further reduce the background, we require the  \xkshort~transverse-decay 
length $L_{K_S^0}$, the distance in the $r-\phi$ plane  between the primary and secondary vertices, 
to be $2.5 < L_{K_S^0} < 50$ cm. The \xlambda~selection criteria are the same as \xkshort~except for the lower limit of 
the decay length requirement, which is 5 cm.  The invariant mass  of the two-track system is 
calculated by  attributing the charged-pion mass to both tracks.  The left plot in Fig.~\ref{fig:mes} 
shows the $\pi^+\pi^-$ invariant mass ($M_{\pi^+\pi^-}$) for \xkshort~candidates with $|\eta|<1$ in MB events.
For the \xlambda~reconstruction, the track with the higher momentum is assigned the proton mass.
Any reference to \xlambda~implies $\overline{\Lambda}^0$ as well. 
The invariant mass distributions are modeled with either a Gaussian or Breit-Wigner function 
for the signal and a third-degree polynomial function for the background. As the widths of particles are small, the 
third-degree polynomial is adequate to model the background shape within the fit range.
 
\section{\xkstar~and \xphi~RECONSTRUCTION}
\begin{figure*}[htp]                                                                                                                                                             
{\psfig{file=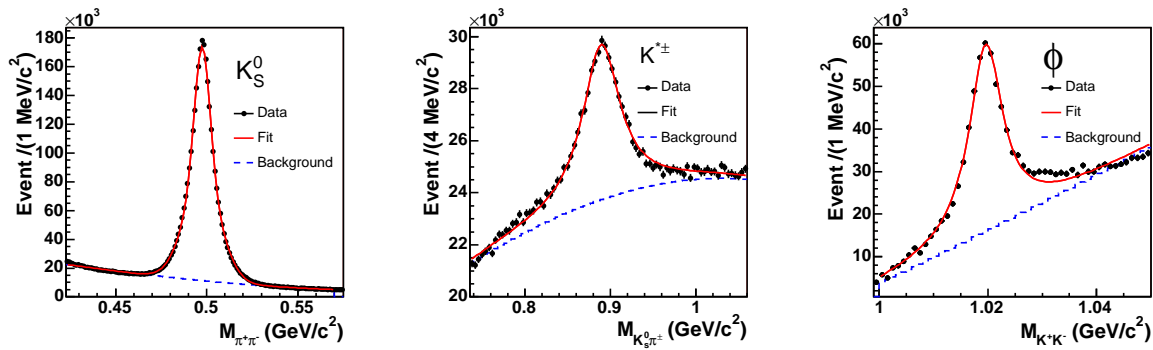,height=5.0cm}}                                                                                                                                       
\caption{Reconstructed invariant mass distributions for charged-pion pairs ($M_{\pi^+\pi^-}$), \xkshort~and 
         pion pairs ($M_{K_{S}^{0}\pi^{\pm}}$)                                                                                           
        and charged-kaon pairs ($M_{K^+K^-}$) from MB events.      
        The solid line is the fitted curve, a third-degree polynomial for the background and a                             
        double Gaussian (\xkshort) or Breit-Wigner (\xkstar~and \xphi) function to model the signal.                               
       The widths are consistent with the mass resolution from the Monte Carlo simulation.}              
\label{fig:mes}                                                    
\end{figure*}

 \xkstar~and \xphi~particles are only reconstructed in MB events. We reconstruct \xkstar~decaying 
 into \xkshort~and \xpi. Since the lifetime of \xkstar~is very short, the reconstructed \xkshort~
 candidates from the previous section with their mass 0.47 $<M_{\pi^+\pi^-}<$ 0.53 GeV/$c^2$  and a track 
 with $p_T>$ 0.325 GeV/$c$ are combined at the event vertex. For both the \xkshort~candidate 
 and the track, we require that the impact parameter $d_0$ to be less than 0.25 cm, and $\delta z_0$ to be 
 less than $2$ cm. The charged-pion mass is assigned to the track. The center plot in Fig.~\ref{fig:mes} shows the 
 invariant mass of a \xkshort~and a charged pion combinations ($M_{K_{S}^{0}\pi^{\pm}}$), and there is a distinct \xkstar~signal.

The final state for \xphi~reconstruction is $K^+$ and  $K^-$. Since the lifetime of \xphi~is also 
very short, two oppositely-charged tracks, assumed to be kaons, with $ p_T>~$0.325 GeV/$c$, are combined at the event vertex 
after requiring $d_0 < 0.25$ cm and $\delta z_0 < 2$ cm for both tracks. The right plot in Fig.~\ref{fig:mes} shows the two-kaon 
invariant mass ($M_{K^+K^-}$) after the same sign $KK$ invariant mass distribution is subtracted. 
There is a mismatch between the data and the fitted curve near $M_{K^+K^-} \sim$~1.03 GeV/$c^2$ 
at the level of a few percentage of the signal events, much less than the systematic uncertainty 
due to the fitting procedure as discussed later.

To measure the $p_T$ cross-section distribution of a resonance, the data in the invariant 
mass distribution are divided into many $p_T$ intervals and the number of resonances  is calculated for 
each $p_T$ interval from a fit to the invariant mass distribution. The numbers as a function of $p_T$ are 
acceptance-corrected to produce the $p_T$ distribution. In this paper, the word resonance is loosely used for both 
short-lived and long-lived particles. 

\section{$p_T$ DISTRIBUTIONS OF $K_S^0$, $K^{\star\pm}$ and $\phi$ MESONS IN MB EVENTS}
\subsection{ACCEPTANCE CALCULATION AND SYSTEMATIC UNCERTAINTIES}
The geometric and kinematic acceptance is estimated with Monte Carlo (MC)  simulations. Each
resonance state is generated with $\sim$~14 fixed $p_T$ values ranging  from 0.5 to 10 
GeV/$c$ and uniform in rapidity for $|y|<2$. A generated  resonance is combined with either one or four non-diffractive 
inelastic MB events generated with the {\sc pythia}  generator. Although the average number  of 
interactions in our data sample is a little less than two, the default acceptance is calculated from the MC 
sample with four MB events and the difference of the acceptance values between the two samples is taken as a
systematic uncertainty. This is because {\sc pythia} underestimates the average event multiplicity.

\begin{figure}[htp]
{\psfig{file=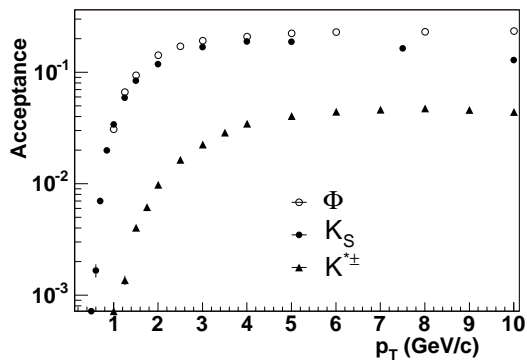,height=5.2cm}}
\caption{Acceptance as a function of $p_T$ for \xkshort, \xkstar~and \xphi~mesons in MB events.
       The values include the branching ratios to the final states detected. }
\label{fig:accmin}
\end{figure}

 The detector response to particles produced in the event generator is modeled with the CDF II detector simulation 
 based on the {\sc geant-3} MC program~\cite{GEANT}. Simulated events are processed and selected 
 with the same analysis code as that used for the data. The acceptance is defined as the ratio of the  number of 
 reconstructed resonances with the input $p_T$ divided by the generated number, including the branching ratio.  
 Acceptance values are calculated separately for the particles and their corresponding antiparticles and the 
 average of the two is used as the  default value, since the acceptances for the two states are similar. 
 Figure~\ref{fig:accmin} shows the acceptances including the relevant branching ratios for the three particles. 

The acceptance values as a function of $p_T$ are  fitted with a fourth-degree polynomial function and the fitted curve 
is used to correct the numbers of each resonance state in the data.  The modeling of the MB
events overlapping with the resonance, and the selection criteria applied, contribute to the systematic 
uncertainty on the acceptance calculation. 
Acceptance uncertainties due to the selection criteria are studied by changing the selection values of the 
variables used to reconstruct the resonances. The variables  examined are $p_T$, $|z_1 - z_2|$, 
$\delta z_0$, $d_0$, and the decay lengths. For each variable other than $p_T$, two values around the default 
value are typically chosen. One value is such that it has little effect on the signal, and the other 
reduces the  signal by approximately 20 to 30$\%$. The default minimum $p_T$ selection value is 0.325 GeV$/c$, 
which is changed to 0.3 GeV$/c$ and to 0.35 GeV$/c$. 

 For each considered variation, a new acceptance curve and number of resonances as a function of $p_T$ are 
 obtained, and the percentage change between the new $p_T$ distribution and that with the default selection 
 requirements is taken as the uncertainty  in the acceptance for the specific $p_T$ interval. The sum in quadrature
 of all variations is taken as the  total uncertainty 
 on the acceptance in a given $p_T$ bin. For the \xkshort~case, the acceptance uncertainty  decreases from 
 about $15\%$ at $p_T \sim~$1 GeV/$c$ to $4\%$ at $p_T \sim 5$ GeV/$c$ and then rises again to $10\%$ 
 at $p_T=10$ GeV/$c$.  This acceptance uncertainty is  added quadratically to the systematic uncertainty due 
 to the fitting procedure, described later, to give the total systematic uncertainty.

For \xkstar~and \xphi~mesons the examined variables are  $p_T$, $\delta z_0$ and  $d_0$ as they decay at the event vertex. 
The acceptance uncertainty for the \xkstar~case decreases from about 25\% at $p_T\sim~1.5$ GeV$/c$ to 10\%  at 
$p_T\sim 5$ GeV$/c$ and then rises to $\sim~15\%$ at 10 GeV/$c$. For the \xphi~meson, the uncertainty decreases 
from about 15\% at $p_T\sim~1$ GeV$/c$ to 10\%  at $p_T\sim~2$ GeV$/c$, decreases to 6\% at $p_T\sim~5$ 
GeV$/c$ and is then constant. 

\subsection{$p_T$ DISTRIBUTIONS}
The first step to get the $p_T$ distribution is to calculate the number of resonances as a function of $p_T$
from the invariant mass plots. The data in the invariant mass plot for each resonance are divided into many $p_T$ 
intervals. The number of $p_T$ intervals depends on the resonance type and is dictated by statistics such that 
the fits to the invariant mass distributions are stable. The number of resonances in each $p_T$ interval is 
determined by fitting the invariant mass distributions using a Gaussian (\xkshort) or non-relativistic Breit-Wigner (\xkstar~
and \xphi) function with three parameters for the signal, and a third-degree polynomial for the  underlying 
background. The measured mass distributions of the \xkstar~and \xphi~are not exactly a Breit-Wigner 
shape because of the detector resolution. The detector effect on the mass shape is treated as one of 
the systematic uncertainties. The polynomial fit to the background is subtracted bin-by-bin from the data 
in the mass interval to obtain  the number of  resonances. This number is divided by the acceptance 
to obtain the  $p_T$ cross-section distributions. Table~\ref{tab:win} shows the mass intervals for each resonance.

The fitting procedure is one source of systematic uncertainty. This uncertainty  is estimated by separately 
varying the mass range of the fit, the functional form for the signal to a double Gaussian function (\xkshort) 
or a Breit-Wigner function convoluted with a Gaussian (\xkstar~and \xphi), and the background modeling function 
to a second-order polynomial. The mass and width of the Breit-Wigner function are fixed to the values in 
the Review of Particle Properties~\cite{ref:pdg}. The number of signal events
is recalculated in all $p_T$ intervals for each variation.  The systematic uncertainty is determined as the 
sum in quadrature  of the fractional change in the number of signal events from each modified fit. Because 
the \xkshort~signals are clearly visible, the systematic uncertainty is low,  less than 5$\%$ up to $p_T$ = 10 GeV/$c$.  
For the \xkstar~case it decreases from about 25\% at $p_T\sim~1$ GeV$/c$ to 6\%  at $p_T\sim 4$ GeV$/c$ and 
then rises to $\sim~10\% $ at 10 GeV/$c$.  For the \xphi~meson, the uncertainty decreases from about 25\% at 
$p_T\sim~1$ GeV$/c$ to 8\%  for $p_T > 2$ GeV$/c$ and remains fairly constant. The high uncertainty 
in the low $p_T$ region is due to a large combinatorial background. 
The total systematic uncertainty is the square root of the quadratic 
sum of the fitting uncertainty in this section and the uncertainty in the acceptance calculation. 

\begin{table}[htp]
\begin{center}
\caption{\label{tab:win} The mass intervals used to select the signal events. 
         The polynomial fit to the background is subtracted bin-by-bin from the data 
         in the mass interval to obtain  the number of  signal events. The unit is GeV/$c^2$.}
\begin{tabular}{ccc}
\hline\hline\\[-1.0em]
 Particle type   & MB events   & Jet events \\
\hline\\[-1.0em]
$\Lambda^0$   &         -                   &  1.105 -- 1.132 \\
$K_S^0$          &    0.48 -- 0.516     &  0.465 -- 0.535 \\
\xkstar             &    0.841 -- 0.943   &   -    \\
\xphi                &    1.01 -- 1.03       &   -    \\
\hline\hline                                                 
\end{tabular}
\end{center}
\end{table}

\begin{figure}[htp]
{\psfig{file=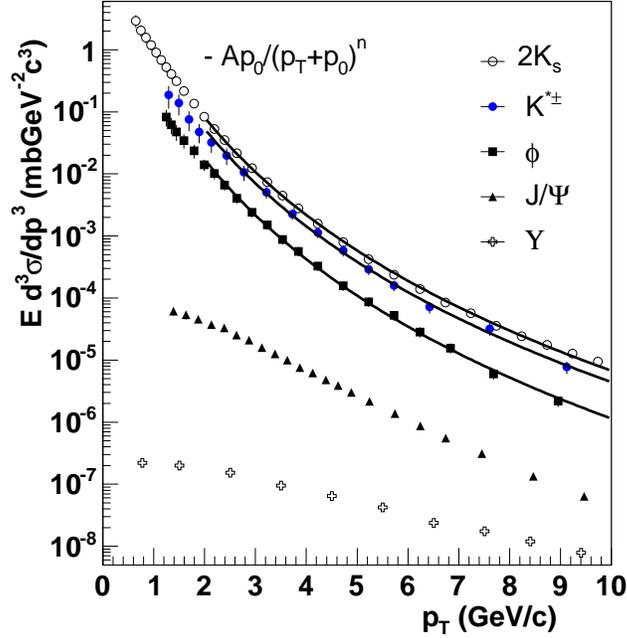,height=10.0cm}}
\caption{The inclusive invariant $p_T$ differential
	cross section distributions ($Ed^3\sigma/dp^3$) for $K_S^0$, $K^{*\pm}$, and
	$\phi$ within $|\eta|<1$. The \xkshort~cross section is multiplied by two to take 
       $K_L^0$ production into account. The solid curves are from fits
	to a power law function, with the fitted parameters given in
	Table \ref{tab:fitmesonp}. The $J/\Psi$~\cite{ref:jpsi} and 
       $\Upsilon$~\cite{ref:upsilon} data are shown for comparison.} 
\label{fig:invmes}
\end{figure}

\begin{figure}[htp]
{\psfig{file=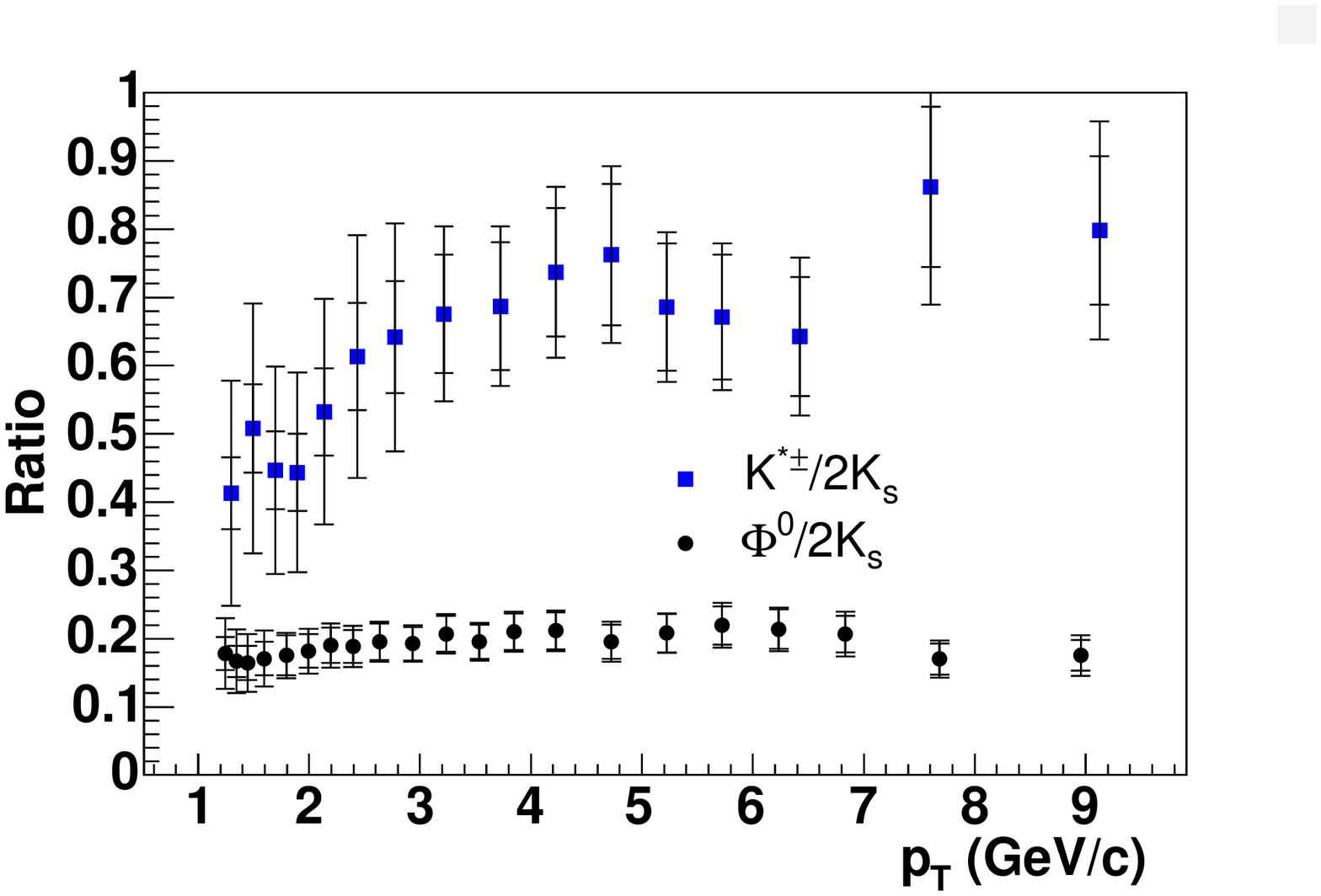,height=5.5cm}}
\caption{The cross-section ratios as a function of $p_T$ of $K^{*\pm}$ to $K_S^0$ and $\phi$ to $K_S^0$.
         The  $K_S^0$ cross section is multiplied by 2.
         There are two error bars for each data point. The inner (outer) one corresponds to the statistical (systematic) uncertainty.}
\label{fig:ratiomes}
\end{figure}

\begin{figure}[htp]
{\psfig{file=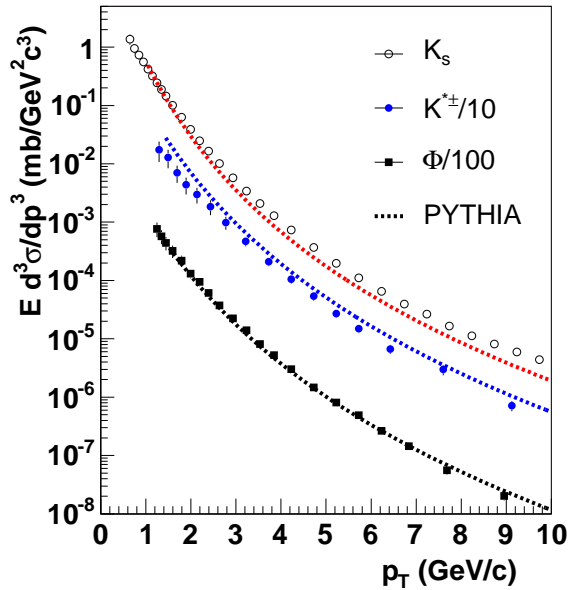,height=9.0cm}}
\caption{The inclusive invariant $p_T$ differential cross section distributions in Fig.~\ref{fig:invmes} are 
         compared with \textsc{pythia} version 6 with default parameters. The \xkstar~(\xphi)~cross section is 
        divided by 10 (100).} 
\label{fig:pythiamb}
\end{figure}

 The inclusive invariant differential cross section as a function of $p_T$ for each particle within $|\eta|<1$ 
 is calculated as $Ed^3\sigma/dp^3$ = $(\sigma_{\rm mb}/N_{\rm event}) d^3N/A p_T dp_T dy d\phi$  
 = $(\sigma_{\rm mb}/2\pi N_{\rm event}) \Delta N/A p_T  \Delta p_T \Delta y $  where $\sigma_{\rm mb}$ 
 is the MB cross section $45\pm 8$ mb~\cite{ref:ndxsection} passing our trigger requirement,  
 $N_{\rm event}$ is the number of events, $\Delta N$ 
 is  the number of resonances observed in each $p_T$ interval ($\Delta p_T$) after the background subtraction, $A$ is the 
 acceptance in the specific $p_T$ interval, and $\Delta y$ is the rapidity range used in the acceptance 
 calculation (-2 $< y <$ 2).

 Figure~\ref{fig:invmes} shows the results for the differential cross sections as a function of $p_T$ for the three resonances. 
 The uncertainties  shown for each data point include the statistical and all systematic uncertainties 
 described above, except for that associated with $\sigma_{\rm mb}$~\cite{ref:ndxsection}. 
 The systematic uncertainties of data points neighboring a $p_T$ value are correlated because the decay kinematics 
 of the daughter particles are similar. The cross sections  in Fig.~\ref{fig:invmes} are listed in Table~\ref{tab:mescross}. 
 The displayed $p_T$ values are the weighted 
 averages within the $p_T$ intervals based on the cross section calculated  from the fit parameters described 
 below. 
 
 The $p_T$ differential cross section is modeled by a power law function, $A(p_0)^n/(p_T + p_0)^n$, for $p_T > 2$ 
 GeV$/c$. In order to compare with the previous publications on hyperons ($\Lambda,~\Xi,~\rm and~\Omega$)~\cite{ref:cdfhyp}, 
 $p_0$ is fixed at 1.3 GeV$/c$, and the results are shown in Table \ref{tab:fitmesonp}.  Compared to hyperons, 
 the values of the parameter $n$ for mesons are lower by $\sim 10\%$.
The data below $p_T\sim~$2 GeV$/c$ cannot be described well by the power law function even if $p_0$ is allowed to 
float. For this region, the data are better described by an exponential function, $Be^{-b \cdot p_T }$. The $p_T$ ranges 
and results of this fit are shown in Table~\ref{tab:fitmesone}, and the slope $b$ of \xphi~is consistent with a previous 
measurement~\cite{ref:e735phi}.  The $b$ values depend on the range of the fit. 
 
 Figure~\ref{fig:ratiomes} shows the  $p_T$ differential cross section ratios of \xkstar~to \xkshort~ 
 and \xphi~to \xkshort. The \xkshort~ cross section is multiplied by two to account for the $K_L^0$  production.
 The \xkstar~to \xkshort~ratio increases as $p_T$ increases,  reaches a plateau at $\sim~5$ GeV/$c$  
 $p_T$ and stays flat. The rise in the \xphi~to \xkshort~ratio at low $p_T$ is slower than the \xkstar~to \xkshort~ratio 
 and reaches the plateau at an earlier $p_T$. The two ratios as a function of $p_T$ exhibit a similar behavior 
 as  $\Xi^\pm$/(\xlambda + \xlambdab) and  $\Omega^\pm$/(\xlambda +~\xlambdab)~\cite{ref:cdfhyp}. 
 
In Figure~\ref{fig:pythiamb}, the differential cross sections of the three resonances are compared with \textsc{pythia} 
events generated with default parameters. The \xphi~cross section matches well while \textsc{pythia} \xkshort~(\xkstar) cross section 
is somewhat lower (higher) than the data. The \textsc{pythia}~parameters responsible for the strange meson production cross 
sections were varied~\cite{ref:private} but it was not possible to produce a good match for all three resonances.

\begin{table*}
\begin{center}
\caption{\label{tab:mescross}   The inclusive invariant differential cross section values for 
                                  \xkshort, \xphi, and \xkstar~mesons in  Fig.~\ref{fig:invmes}.
                                  The uncertainties include  both the statistical and systematic uncertainties 
                                   added in quadrature but do not include $\sigma_{\rm mb}$ uncertainty.}
\begin{tabular}{cc|cc|cc}\hline\hline\\[-1.0em]
$p_T$ (GeV/$c$) & 2\xkshort~($mb~\rm GeV^{-2}c^3$) & ~~$p_T~~$  & \xphi & $~~p_T~~$  & \xkstar\\
\hline\\[-1.0em]
.645  & $ 2.94\pm0.66 $                                 & 1.24  & $(8.25\pm2.48)$\x$10^{-2}$          & 1.29  & $(1.87\pm0.74)$\x$10^{-1}$\\
.745  & $ 2.05\pm0.43 $                                 & 1.34  & $(6.17\pm1.80)$\x$10^{-2}$          & 1.49  & $(1.39\pm0.51)$\x$10^{-1}$\\
.845  & $ 1.58\pm0.30 $                                 & 1.44  & $(4.75\pm1.34)$\x$10^{-2}$          & 1.69  & $(7.56\pm2.63)$\x$10^{-2}$ \\
.945  & $1.19\pm0.22$                                   & 1.59  & $(3.45\pm0.91)$\x$10^{-2}$          & 1.90  & $(4.72\pm1.60)$\x$10^{-2 }$\\
1.04  & $(9.09\pm1.58)$\x$10^{-1}$     & 1.79  & $(2.36\pm0.58)$\x$10^{-2}$           & 2.13  & $(3.21\pm1.02)$\x$10^{-2}$ \\
1.15  & $(6.88\pm1.15)$\x$10^{-1}$     & 1.99  & $(1.41\pm0.29)$\x$10^{-2}$           & 2.44  & $(1.98\pm0.60)$\x$10^{-2}$ \\
1.24  & $(5.25\pm0.84)$\x$10^{-1}$      & 2.19  & $(1.01\pm0.20)$\x$10^{-2}$          & 2.77  & $(1.06\pm0.29)$\x$10^{-2}$ \\
1.34  & $(4.06\pm0.63)$\x$10^{-1}$     & 2.40  & $(6.61\pm1.23)$\x$10^{-3}$           & 3.21  & $(5.06\pm1.07)$\x$10^{-3}$ \\
1.44  & $(3.13\pm0.48)$\x$10^{-1}$     & 2.63  & $(4.03\pm0.72)$\x$10^{-3}$           & 3.73  & $(2.25\pm0.44)$\x$10^{-3}$ \\
1.59  & $(2.17\pm0.32)$\x$10^{-1}$     & 2.94  & $(2.40\pm0.41)$\x$10^{-3}$           & 4.22  & $(1.13\pm0.22)$\x$10^{-3}$ \\
1.79  & $(1.35\pm0.19)$\x$10^{-1}$     & 3.23  & $(1.51\pm0.26)$\x$10^{-3}$           & 4.72  & $(5.80\pm1.13)$\x$10^{-4}$ \\
1.99  & $(8.37\pm1.14)$\x$10^{-2}$     & 3.54  & $(8.73\pm1.49)$\x$10^{-4}$           & 5.22  & $(2.90\pm0.56)$\x$10^{-4}$ \\
2.19  & $(5.32\pm0.71)$\x$10^{-2}$    & 3.84  &  $(5.67\pm0.97)$\x$10^{-4}$           & 5.72  & $(1.60\pm0.31)$\x$10^{-4}$ \\
2.40  & $(3.53\pm0.46)$\x$10^{-2}$    & 4.22  &  $(3.26\pm0.56)$\x$10^{-4}$           & 6.42  & $(7.16\pm1.51)$\x$10^{-5}$ \\
2.63  & $(2.17\pm0.28)$\x$10^{-2}$    & 4.72  &  $(1.57\pm0.27)$\x$10^{-4}$           & 7.61  & $(3.21\pm0.73)$\x$10^{-5}$ \\
2.94  & $(1.24\pm0.16)$\x$10^{-2}$    & 5.22  &  $(8.69\pm1.55)$\x$10^{-5}$           & 9.13  & $(7.70\pm1.76)$\x$10^{-6}$ \\
3.23  & $(7.33\pm0.96)$\x$10^{-3}$    & 5.72  &  $(5.25\pm0.94)$\x$10^{-5}$ &  - &  - \\ 
3.54  & $(4.50\pm0.59)$\x$10^{-3}$    & 6.24  &  $(2.84\pm0.53)$\x$10^{-5}$ &  - &  - \\ 
3.84  & $(2.78\pm0.36)$\x$10^{-3}$    & 6.83  &  $(1.56\pm0.29)$\x$10^{-5}$ &  - &  - \\
4.22  & $(1.57\pm0.20)$\x$10^{-3}$    & 7.68  &  $(6.01\pm1.17)$\x$10^{-6}$ &  - &  - \\ 
4.72  & $(7.91\pm1.03)$\x$10^{-4}$    & 8.96  &  $(2.18\pm0.42)$\x$10^{-6}$ &  - &  - \\
5.22  & $(4.22\pm0.55)$\x$10^{-4}$ &  - &  - &  - &  - \\ 
5.72  & $(2.38\pm0.31)$\x$10^{-4}$ &  - &  - &  - &  - \\ 
6.24  & $(1.39\pm0.18)$\x$10^{-4}$ &  - &  - &  - &  - \\ 
6.74  & $(8.46\pm1.10)$\x$10^{-5}$ &  - &  - &  - &  - \\ 
7.24  & $(5.70\pm0.74)$\x$10^{-5}$ &  - &  - &  - &  - \\ 
7.74  & $(3.57\pm0.47)$\x$10^{-5}$ &  - &  - &  - &  - \\ 
8.24  & $(2.43\pm0.32)$\x$10^{-5}$ &  - &  - &  - &  - \\ 
8.74  & $(1.75\pm0.23)$\x$10^{-5}$ &  - &  - &  - &  - \\ 
9.24  & $(1.28\pm0.17)$\x$10^{-5}$ &  - &  - &  - &  - \\ 
9.74  & $(9.42\pm1.23)$\x$10^{-6}$ &  - &  - &  - &  - \\
\hline \hline
\end{tabular}
\end{center}
\end{table*}

\begin{table*}
\begin{center}
\caption{\label{tab:fitmesonp} The results of power law function fits to the inclusive
        invariant $p_T$ differential cross sections shown in Fig.~\ref{fig:invmes}
        for $p_T>2$ GeV$/c$.  The parameter $p_0$ is fixed to $1.3$ GeV$/c$ in
        all fits. \xkshort~results are without the scale factor two which takes into account  the $K_L^0$ meson.
       The $K_S^0$ values in the second column are from $\sqrt{s}=1.8$
        TeV~\cite{ref:cdfks}. The uncertainties do not include $\sigma_{\rm mb}$
        uncertainty.  The last line of the table gives the $\chi^2$
        per degree-of-freedom of the fit to data.}
\begin{tabular}{ccccc}
\hline\hline\\[-1.0em]
Fit parameter (units) & $K_S^0$~[Run I] & $K_S^0$  & $K^{*\pm}$ & $\phi$ \\
\hline\\[-1.0em]
$A$ ($mb$ GeV$^{-2}c^{3}$) & $45\pm9$    & $50.2\pm6.1$ & $60.4\pm13.5$  & $23.5\pm2.55$ \\
$p_0$ (GeV/$c$)                     & $1.3$           & $1.3$              & $1.3$                 & $1.3$         \\
$n$                                     & $7.7\pm0.2$ & $7.65\pm0.08$ & $7.60\pm0.19$  & $7.80\pm0.80$ \\
$\chi^2$/dof                       & $8.1/11$      & $6.0/17$        & $3.9/10$            & $14.0/13$     \\
\hline\hline
\end{tabular}
\end{center}
\end{table*}

\begin{table*}
\begin{center}
\caption{\label{tab:fitmesone} The results of exponential function fits to the inclusive
        invariant $p_T$ differential cross sections shown in Fig.~\ref{fig:invmes}
        for the $p_T$ ranges given in the second row.  
       The \xkshort~results are without the scale factor two which takes into account the $K_L^0$ meson.
        The uncertainties shown do not include $\sigma_{\rm mb}$ uncertainty.
        The last line of the table gives the $\chi^2$ per degree-of-freedom of
        the fit to data.}
\begin{tabular}{cccccc}\hline\hline\\[-1.0em]
Fit parameter (units) & $K_S^0$    & $K_S^0$   &  $K_S^0$   & $K^{*\pm}$ & $\phi$ \\
$p_T$ range (GeV/$c$) & [0.6, 1.5] & [0.6, 2.5] & [1.2, 2.5]& [1.2, 2.5] & [1.2, 2.5] \\
\hline\\[-1.0em]
$B$ ($mb$ GeV$^{-2}c^{3}$) & $10.4\pm2.4$  & $6.55\pm0.80$ & $5.00\pm0.92$  & $1.79\pm1.203$ & $1.20\pm0.40$ \\
$b$ (GeV$^{-1}c$)       & $3.02\pm0.20$ & $2.60\pm0.08$ & $2.41\pm0.10$  & $2.01\pm0.35$  & $2.20\pm0.18$ \\
$\chi^2$/dof              & $1.0/9$       & $7.2/12$      & $0.7/6$      &   $0.8/4$        & $0.3/7$       \\
\hline \hline
\end{tabular}
\end{center}
\end{table*}

\section{$p_T$ DISTRIBUTIONS OF $K_S^0$ AND $\Lambda^{0}$ HADRONS IN JETS}
\subsection{ACCEPTANCE CALCULATION AND SYSTEMATIC UNCERTAINTIES}
\begin{figure}[htp]
{\psfig{file=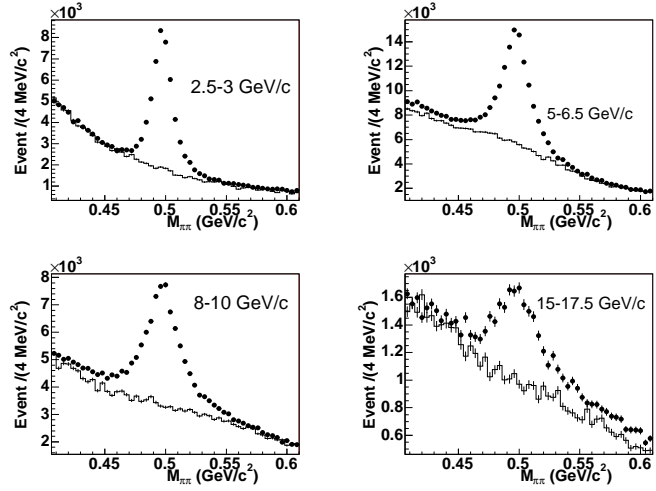,height=6.75cm}}
\caption{Invariant $\pi^+\pi^-$ mass distributions for four $p_T$ intervals  from jets 
         with $E_T$ between 60 and 80 GeV.  The numbers in the figures are the $p_T$ intervals.
        The histograms are the background shapes called QCD-C and obtained from QCD jet simulation. }
\label{fig:ksj}
\end{figure}

\begin{figure}[htp]
{\psfig{file=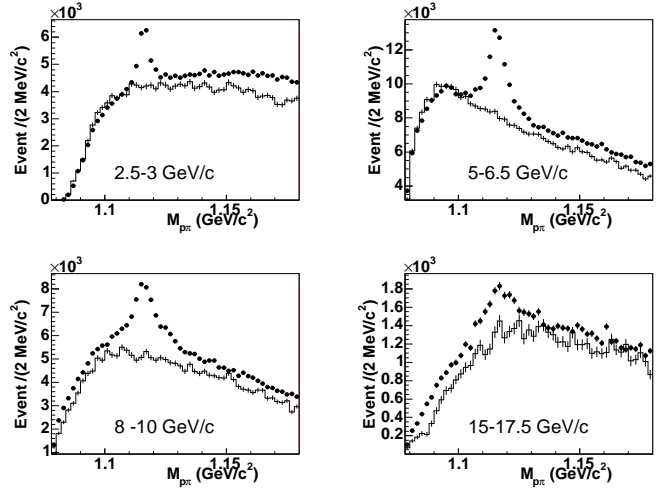,height=6.75cm}}
\caption{Invariant $p\pi^-$ + $\bar{p}\pi^+$ mass distributions for four $p_T$ intervals  from jets 
         with $E_T$ between 60 and 80 GeV.  The numbers in the figures are the $p_T$ intervals.
        The histograms are the background shapes called QCD-C and obtained from QCD jet simulation. 
        Because the \xlambda~signal becomes unclear as $p_T$ increases (bottom right plot), the data
        with $p_T$ greater than 15 GeV/$c$ and  jet $E_T$ greater than 60 GeV are not used.}
\label{fig:lamj}
\end{figure}

For jet events, \xkshort~and \xlambda~ candidates reconstructed as previously discussed are divided into five jet-$E_T$ ranges. 
A candidate is assigned to a jet if $\Delta R<$~0.5, where $\Delta R$ is the distance between the 
resonance and  jet in the $\eta-\phi$ plane. If the candidate belongs to more than one jet, it is associated to the nearest jet. 
The $\Delta R$ range 0.5 is slightly larger than the 0.4 used in the jet clustering to include 
low $p_T$ resonances. Figure~\ref{fig:ksj} shows $M_{\pi^+\pi^-}$ distributions from jets with 60 $<E_T<$ 80 GeV  
and Fig.~\ref{fig:lamj} shows the same but for the $M_{p\pi^-}$~+ $M_{\bar{p}\pi^+}$~distributions.  
Because at large $p_T$ the \xlambda~signal becomes unclear (bottom right plot in  Fig.~\ref{fig:lamj}), the \xlambda~
data with $p_T >$ 15 GeV/$c$  and jet $E_T >$ 60 GeV are not used.

The acceptance for \xkshort~and \xlambda~hadrons in jets as a function of $p_T$ is calculated for each jet-$E_T$ interval 
and defined as the ratio of the number of reconstructed resonances to the number of generated resonances in 
the jets.  The acceptances in jets are calculated using the QCD  jet events generated with  {\sc pythia}, passed
through the CDF II detector simulation, and reconstructed. A jet event is mixed with one or four  
{\sc pythia} inelastic MB events. The default acceptance is calculated with the sample mixed with
four MB events, and the difference of the acceptance values between the two samples 
is one of  our systematic uncertainties, as in the case of MB events. 

\begin{figure}[htp]
{\psfig{file=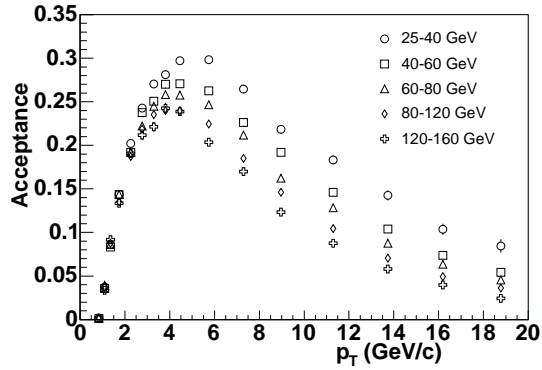,height=5.2cm}}
\caption{\xkshort~acceptance plots for the five jet-$E_T$ intervals. 
       The values include the branching ratio $K^0_s \rightarrow \pi^+\pi^-$. }
\label{fig:accjks}
\end{figure}

\begin{figure}[htp]
{\psfig{file=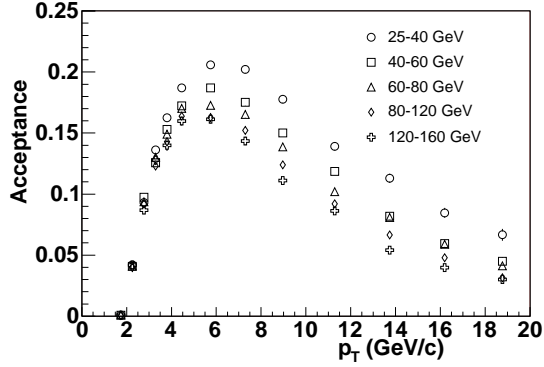,height=5.2cm}}
\caption{ \xlambda~acceptance plot for the five jet-$E_T$ intervals. 
       The values include the branching ratio $\Lambda^0 \rightarrow p \pi^-$. }
\label{fig:accjlam}
\end{figure}

We select the generated resonances in the MC data with $\Delta R <$~0.5 where $\Delta R$ is measured with 
respect to the reconstructed jet direction.  We also select the reconstructed resonances within the same $\Delta R$ 
range, and mark the ones with matched generated resonances based on  
$|\Delta\eta|<$ 0.075 and $|\Delta\phi|<$ 0.075, where $\Delta\phi$ ($\Delta\eta$) is the difference in 
$\phi$ ($\eta$) between the generated and reconstructed resonances. The acceptance as a function of $p_T$ 
is the ratio of the $p_T$ distribution of the marked reconstructed resonances to the generated 
resonances. Figure~\ref{fig:accjks} shows the \xkshort~acceptance for the five jet-$E_T$ intervals and  
Fig.~\ref{fig:accjlam} shows the same for \xlambda. The acceptances include the branching ratio to our final states.

The sources of systematic uncertainty in the acceptance calculation are similar to those discussed for 
\xkshort~in MB events, and they are calculated as functions of $p_T$ and $E_T$ except 
for one difference: The default minimum $p_T$ selection 0.5 GeV$/c$ is changed to 0.45 GeV$/c$ 
and to 0.55 GeV$/c$. The dependence of the \xkshort~(\xlambda) acceptance uncertainty on $p_T$ for 
different $E_T$ ranges is quite similar. It starts at $\sim 10~(15)\%$ at 2 GeV/$c$ and decreases to  
$\sim 5~(7)\%$ at 5 GeV/$c$ and then increases approximately linearly to $\sim 12~(20)\%$ at 20 GeV/$c$.

\subsection{$p_T$ DISTRIBUTIONS}
The measurement of the $p_T$ distribution of particles in jets is different from that in MB events  because 
there is more combinatorial background. We subtract the background obtained from the simulated QCD MC data 
sample before fitting the mass distribution. The background is called QCD combinatorial (QCD-C) background and it  
is the $M_{\pi\pi}$ (or $M_{p\pi}$) distribution without \xkshort~(or \xlambda). 

\begin{figure}[htp]
{\psfig{file=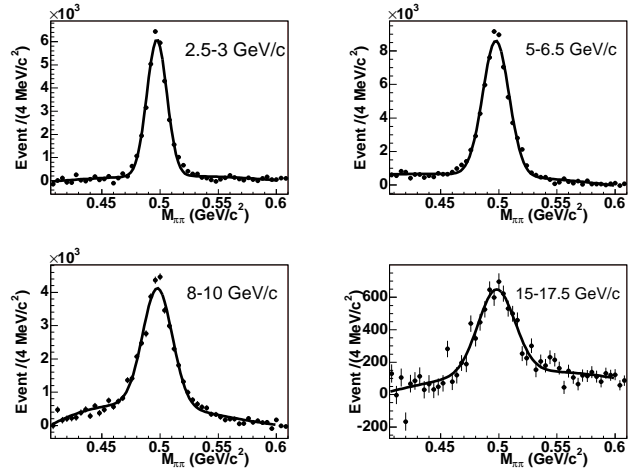,height=6.5cm}}
\caption{Invariant $\pi^+\pi^-$ mass distributions after subtracting the scaled QCD-C background histogram from data
          in Fig.~\ref{fig:ksj}. There are four $p_T$ intervals  and the $E_T$ of jets is 
          between 60 and 80 GeV. The solid lines are fitted curves, a third-degree polynomial 
          for the background and a Gaussian function with three parameters to model the \xkshort~signal. }
\label{fig:ksjsub}
\end{figure}

\begin{figure}[htp]
{\psfig{file=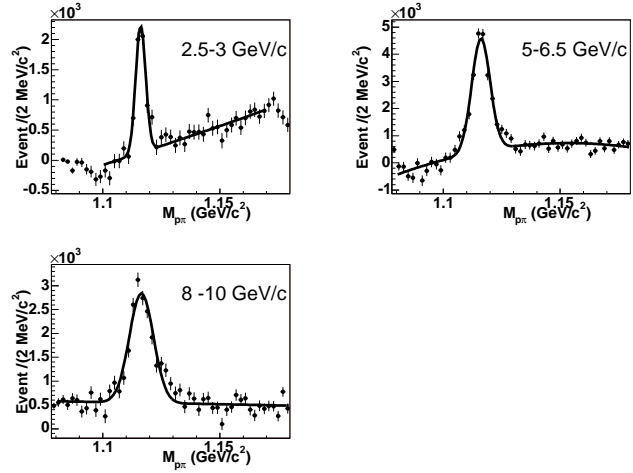,height=6.5cm}}
\caption{ Invariant $p\pi^-$ + $\bar{p}\pi^+$ mass distributions after subtracting the scaled QCD-C background 
          histogram from data in Fig.~\ref{fig:lamj}. There are three $p_T$ intervals  and $E_T$ of jets is 
          between 60 and 80 GeV. The solid lines are fitted curves, a third-degree polynomial 
          for the background and a Gaussian function with three parameters to model the \xlambda~signal.}
\label{fig:lamjsub}
\end{figure}

The QCD-C background shape is obtained as follows. After choosing two tracks that form a \xkshort~(or \xlambda) candidate, 
we check if the candidate has a corresponding \xkshort~(or \xlambda) at the MC particle generation level in the same event by 
comparing the kinematic variables ($\phi$ and $\eta$). If the candidate has a 
corresponding particle at the generation level, the candidate is not entered in the invariant mass
distributions and the distributions are the QCD-C backgrounds shown in Figs.~\ref{fig:ksj} and 
~\ref{fig:lamj}. Disagreement with the data outside the resonance mass regions 
is expected since the shape of the invariant mass distribution is sensitive to the  particle multiplicity and kinematics 
from jets. Figures~\ref{fig:ksjsub} and~\ref{fig:lamjsub} show the invariant mass distributions after 
subtracting the QCD-C backgrounds, scaled such that the entries are mostly positive after subtraction. 
The effect of the normalization is one of the systematic uncertainties.

The number of signal events in each $p_T$ interval is determined by fitting the 
background-subtracted invariant mass distributions using a Gaussian function for the signal and a third-degree
polynomial for the remaining background. The curves from the fits are displayed in the same figures. 
The polynomial function representing the background is subtracted bin-by-bin from the data  
in the mass interval to obtain  the number of  signal events. Table~\ref{tab:win} shows the mass intervals. 
The mass intervals for jets are wider because the $p_T$ range is extended to 20 GeV/$c$ and mass resolution 
gets worse (see Figs.~\ref{fig:ksjsub} and~\ref{fig:lamjsub}) as $p_T$ increases.

The QCD-C background subtraction and  the fitting procedure are 
sources of systematic uncertainty.  The estimation of the fitting procedure uncertainty is similar 
to that for \xkshort~in MB events. The uncertainty from the QCD-C background subtraction 
is estimated by scaling the background by $-25\%$ from the default and recalculating the number of 
signal events. Similar to the \xkshort~in the MB events, the \xkshort~signal is clearly visible
and the uncertainty is fairly constant at $\sim~12\%$ for all $p_T$ and $E_T$  intervals.  For the \xlambda~baryon, 
the uncertainty increases by  $\sim~2\%$ at the high $p_T$ and   $E_T$ region, and we assign a conservative 
17\% for all $p_T$ and $E_T$  intervals. The total systematic uncertainty is the 
uncertainty discussed above and the uncertainty in the acceptance calculation added in quadrature.

\begin{figure}[htp]
{\psfig{file=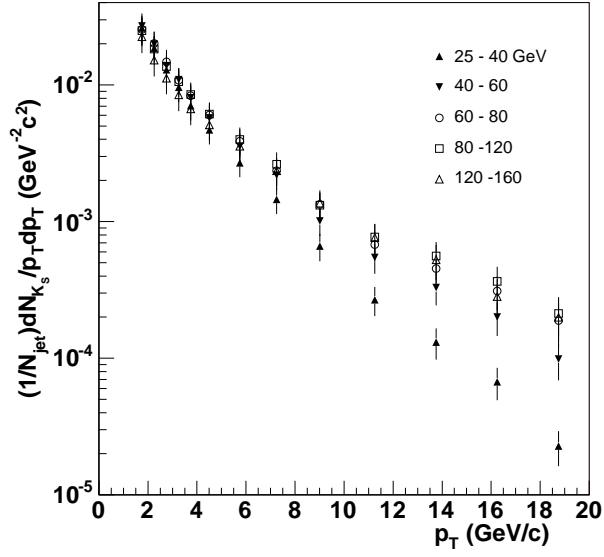,height=8.5cm}}
\caption{The $p_T$ distributions of \xkshort~mesons in centrally-produced jets ($|\eta|<1$) for the five jet-$E_T$ intervals.
                                   The uncertainties include  both the statistical and systematic uncertainties added in quadrature.}
\label{fig:ptksj}
\end{figure}

\begin{figure}[htp]
{\psfig{file=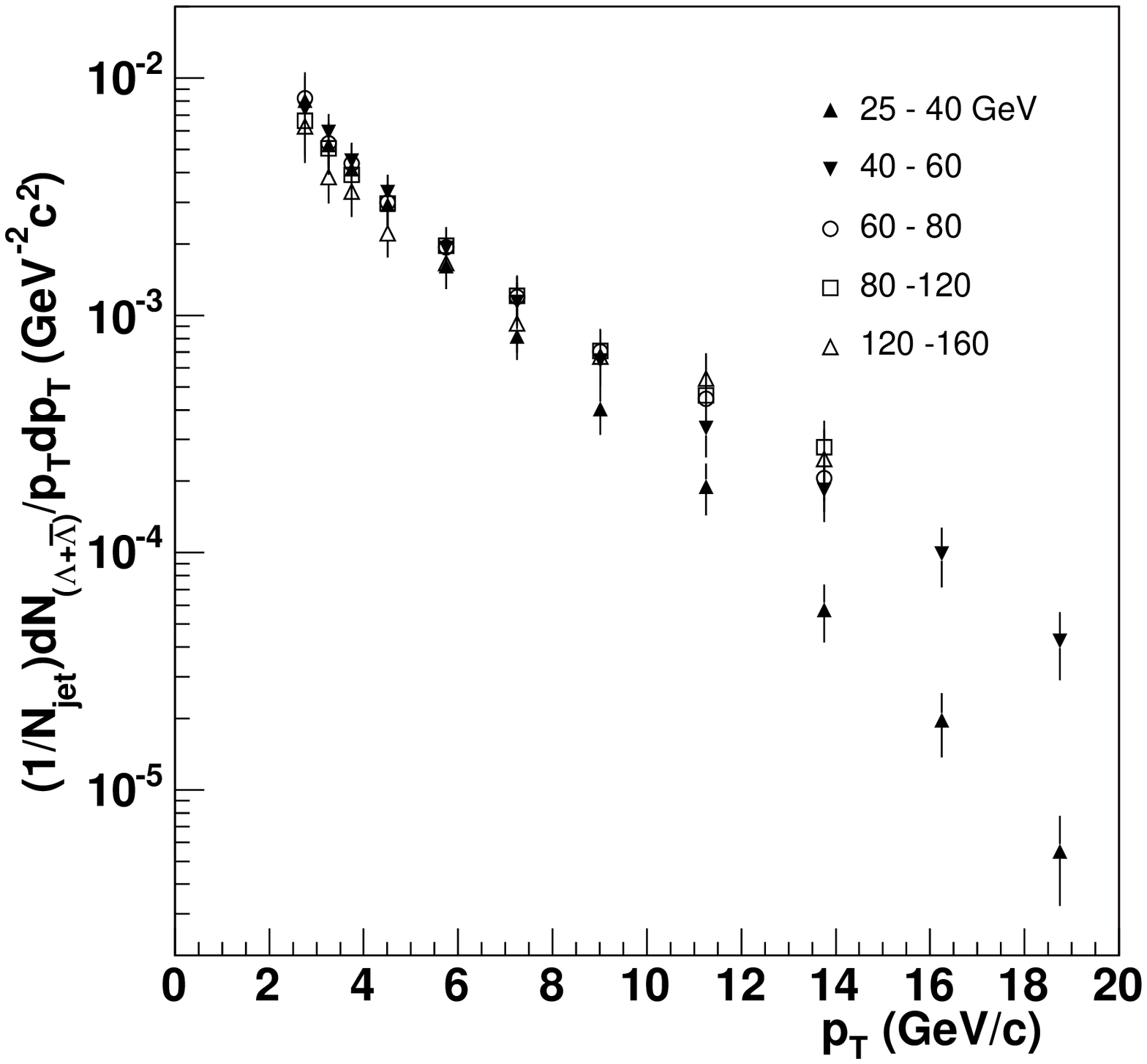,height=8.5cm}}
\caption{The $p_T$ distributions of \xlambda~+ \xlambdab~baryons in centrally-produced jets ($|\eta|<1$) for the five jet-$E_T$ 
     intervals. The uncertainties include  both the statistical and systematic uncertainties added in quadrature.}
\label{fig:ptlamj}
\end{figure}

\begin{figure*}[htp]
{\psfig{file=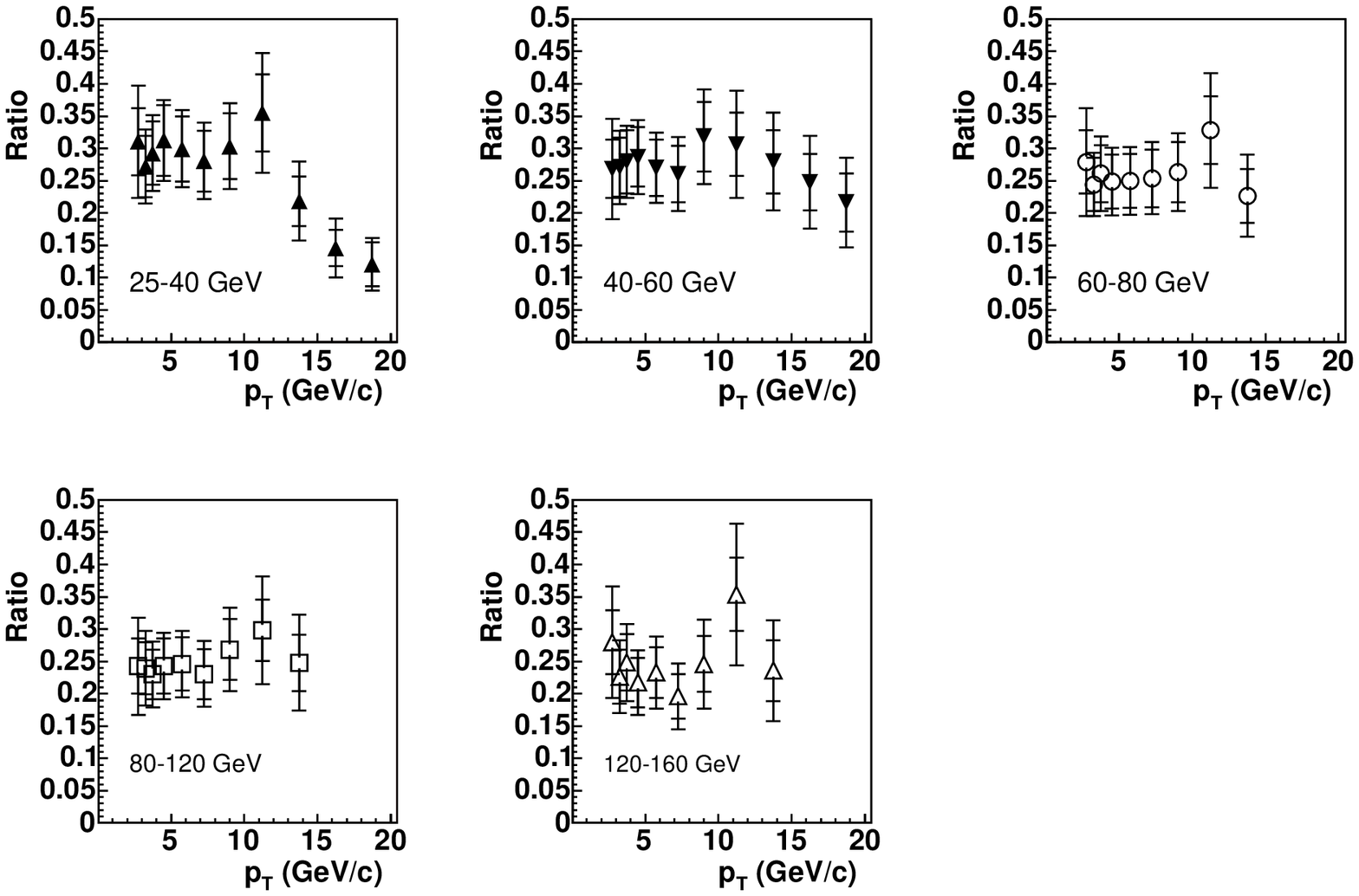,height=9.5cm}}
\caption{The cross-section ratios of \xlambda~+ \xlambdab to 2\xkshort~as a function of $p_T$ for the five jet-$E_T$ intervals.
       There are two error bars for each data point. The inner (outer) one corresponds to the statistical (systematic) uncertainty.}
\label{fig:ratiokl}
\end{figure*}

The $p_T$ distributions are calculated per jet, $1/N_{\rm jet}dN/(p_T dp_T)$ = $1/N_{\rm jet}\Delta N/(A p_T\Delta p_T)$), 
and are shown in  Fig.~\ref{fig:ptksj} (\xkshort) and Fig.~\ref{fig:ptlamj} (\xlambda~+ \xlambdab) for the five jet-$E_T$ 
intervals.  $N_{jet}$ is the number of jets in the $E_T$ interval, $\Delta N$ is the number of signal events  in 
the $p_T$ interval ($\Delta p_T$) and $E_T$ interval, and $A$ is the acceptance at the $p_T$ and $E_T$ interval 
(Figs.~\ref{fig:accjks} and~\ref{fig:accjlam}). The uncertainty is the statistical and systematic uncertainties 
added in quadrature. The average jet-$E_T$ values for the five jet-$E_T$ intervals are 31, 50, 70, 99 and 136 GeV.
Figure~\ref{fig:ratiokl} shows \xlambda~+ \xlambdab to 2\xkshort~ratios as a function of $p_T$ calculated from Figs.~\ref{fig:ptksj} 
and~\ref{fig:ptlamj}. The \xkshort~cross section is multiplied by two to take into account the $K_L^0$ 
production.  The ratios are about 0.25 for all $p_T$ and $E_T$ intervals.

\begin{table*}
\begin{center}
\caption{\label{tab:jetkscross}  The \xkshort~cross-section values for the five jet-$E_T$ intervals 
                                                                in Fig.~\ref{fig:ptksj} }
\begin{tabular}{c|c|c|c|c|c}\hline\hline\\[-1.0em]
    &  \multicolumn{5}{c}{$1/N_{\rm jet}dN/(p_T dp_T)$ (GeV$^{-2}c^{2}$)}\\
\hline 
  $p_T (\rm GeV/c$)    & $E_T:~$25--40 GeV & 40--60 GeV& 60--80 GeV& 80--120 GeV& 120--160 GeV\\
\hline\\[-1.0em]
1.75  & $ (2.57\pm0.59)$\x$10^{-2}$ & $(2.67\pm0.61)$\x$10^{-2}$ &  $ (2.51\pm0.59)$\x$10^{-2}$ &  $(2.47\pm0.58)$\x$10^{-2}$  &  $(2.22\pm0.53)$\x$10^{-2} $ \\ 
2.25  & $ (1.86\pm0.43)$\x$10^{-2}$ & $(1.99\pm0.45)$\x$10^{-2}$ &  $ (2.02\pm0.46)$\x$10^{-2}$ &  $(1.83\pm0.43)$\x$10^{-2}$  &  $(1.51\pm0.37)$\x$10^{-2} $ \\ 
2.75  & $ (1.29\pm0.29)$\x$10^{-2}$ & $(1.39\pm0.32)$\x$10^{-2}$ &  $ (1.47\pm0.33)$\x$10^{-2}$ &  $(1.37\pm0.31)$\x$10^{-2}$  &  $(1.12\pm0.27)$\x$10^{-2} $ \\ 
3.25  & $ (9.62\pm0.22)$\x$10^{-3}$& $(1.09\pm0.24)$\x$10^{-2}$  &  $ (1.08\pm0.24)$\x$10^{-2}$ &  $(1.05\pm0.24)$\x$10^{-2}$  &  $(8.48\pm2.03)$\x$10^{-3} $ \\ 
3.75  & $ (7.06\pm1.55)$\x$10^{-3}$& $(8.05\pm1.80)$\x$10^{-3}$  &  $ (8.35\pm1.84)$\x$10^{-3}$ &  $(8.52\pm1.88)$\x$10^{-3}$  &  $(6.68\pm1.60)$\x$10^{-3} $ \\ 
4.50  & $ (4.70\pm1.02)$\x$10^{-3}$& $(5.75\pm1.28)$\x$10^{-3}$  &  $ (6.02\pm1.33)$\x$10^{-3}$ &  $(6.11\pm1.35)$\x$10^{-3}$  &  $(5.10\pm1.20)$\x$10^{-3} $ \\ 
5.75  & $( 2.70\pm0.59)$\x$10^{-3}$& $(3.56\pm0.79)$\x$10^{-3}$  &  $ (3.87\pm0.86)$\x$10^{-3}$ &  $(3.99\pm0.88)$\x$10^{-3}$   & $(3.56\pm0.84)$\x$10^{-3} $ \\ 
7.25  & $ (1.45\pm0.32)$\x$10^{-3}$& $(2.19\pm0.48)$\x$10^{-3}$  &  $ (2.36\pm0.52)$\x$10^{-3}$ &  $(2.63\pm0.58)$\x$10^{-3}$   & $(2.36\pm0.56)$\x$10^{-3} $ \\ 
9.01  & $ (6.63\pm1.49)$\x$10^{-4}$& $(1.01\pm0.23)$\x$10^{-3}$  &  $ (1.34\pm0.31)$\x$10^{-3}$ &  $(1.32\pm0.30)$\x$10^{-3}$   & $(1.36\pm0.33)$\x$10^{-3} $ \\ 
11.2  & $ (2.68\pm0.64)$\x$10^{-4}$& $(5.48\pm1.30)$\x$10^{-4}$  &  $ (6.78\pm1.65)$\x$10^{-4}$ &  $(7.72\pm1.88)$\x$10^{-4}$   & $(7.65\pm1.98)$\x$10^{-4} $ \\ 
13.7  & $ (1.31\pm0.34)$\x$10^{-4}$& $(3.29\pm0.85)$\x$10^{-4}$  &  $ (4.54\pm1.21)$\x$10^{-4}$ &  $(5.61\pm1.49)$\x$10^{-4}$   & $(5.25\pm1.50)$\x$10^{-4} $ \\ 
16.2  & $ (6.73\pm1.79)$\x$10^{-5}$& $(2.00\pm0.54)$\x$10^{-4} $ &  $ (3.10\pm0.86)$\x$10^{-4}$ &  $(3.66\pm1.02)$\x$10^{-4}$   & $(2.83\pm0.95)$\x$10^{-4} $ \\ 
18.7  & $( 2.28\pm0.66)$\x$10^{-5}$& $(9.85\pm2.95)$\x$10^{-5} $ &  $ (1.89\pm0.57)$\x$10^{-4}$ &  $(2.12\pm0.66)$\x$10^{-4}$   & $(2.00\pm0.80)$\x$10^{-4} $ \\
\hline \hline
\end{tabular}
\end{center}
\end{table*}

\begin{table*}
\begin{center}
\caption{\label{tab:jetlamcross}   The \xlambda~+ \xlambdab~cross-section values for the five jet-$E_T$ intervals 
                                                                in Fig.~\ref{fig:ptlamj}}
\begin{tabular}{c|c|c|c|c|c}\hline\hline\\[-1.0em]
    &  \multicolumn{5}{c}{$1/N_{\rm jet}dN/(p_T dp_T)$ (GeV$^{-2}c^{2}$)}\\
\hline 
  $p_T (\rm GeV/c$)    & $E_T:~$25--40 GeV & 40--60 GeV& 60--80 GeV& 80--120 GeV& 120--160 GeV\\
\hline\\[-1.0em]
2.75  & $ (8.03\pm2.15)$\x$10^{-3}$& $(7.42\pm2.02)$\x$10^{-3}$& $ (8.24\pm2.35)$\x$10^{-3}$& $(6.62\pm1.96)$\x$10^{-3}$& $(6.27\pm1.87)$\x$10^{-3} $ \\ 
3.25  & $ (5.23\pm1.00)$\x$10^{-3}$& $(5.92\pm1.15)$\x$10^{-3}$& $ (5.30\pm1.16)$\x$10^{-3}$& $(5.20\pm1.13)$\x$10^{-3}$& $(3.84\pm0.89)$\x$10^{-3} $ \\ 
3.75  & $ (4.13\pm0.78)$\x$10^{-3}$& $(4.50\pm0.85)$\x$10^{-3}$& $ (4.36\pm0.91)$\x$10^{-3}$& $(3.92\pm0.82)$\x$10^{-3}$& $(3.32\pm0.73)$\x$10^{-3} $ \\ 
4.50  & $ (2.93\pm0.55)$\x$10^{-3}$& $(3.30\pm0.62)$\x$10^{-3}$& $ (3.00\pm0.60)$\x$10^{-3}$& $(2.97\pm0.59)$\x$10^{-3}$& $(2.22\pm0.47)$\x$10^{-3} $ \\ 
5.75  & $ (1.61\pm0.30)$\x$10^{-3}$& $(1.92\pm0.36)$\x$10^{-3}$& $ (1.94\pm0.38)$\x$10^{-3}$& $(1.96\pm0.39)$\x$10^{-3}$& $(1.66\pm0.36)$\x$10^{-3} $ \\ 
7.25  & $ (8.16\pm1.67)$\x$10^{-4}$& $(1.14\pm0.23)$\x$10^{-3}$& $ (1.20\pm0.26)$\x$10^{-3}$& $(1.21\pm0.26)$\x$10^{-3}$& $(9.24\pm2.29)$\x$10^{-4} $ \\ 
9.01  & $( 4.02\pm0.88)$\x$10^{-4}$& $(6.44\pm1.41)$\x$10^{-4}$& $ (7.03\pm1.61)$\x$10^{-4}$& $(7.08\pm1.68)$\x$10^{-4}$& $(6.70\pm1.78)$\x$10^{-4} $ \\ 
11.2  & $ (1.90\pm0.48)$\x$10^{-4}$& $(3.35\pm0.83)$\x$10^{-4}$& $ (4.45\pm1.11)$\x$10^{-4}$& $(4.60\pm1.23)$\x$10^{-4}$& $(5.41\pm1.53)$\x$10^{-4} $ \\ 
13.7  & $ (5.74\pm1.58)$\x$10^{-5}$& $(1.84\pm0.49)$\x$10^{-4}$& $ (2.06\pm0.57)$\x$10^{-4}$& $(2.78\pm0.82)$\x$10^{-4}$& $(2.47\pm0.82)$\x$10^{-4} $ \\ 
16.2  & $ (1.97\pm0.60)$\x$10^{-5}$& $(9.92\pm2.79)$\x$10^{-5}$& - & - &  -  \\ 
18.7  & $ (5.50\pm2.27)$\x$10^{-6}$& $(4.26\pm1.37)$\x$10^{-5}$& - & - &  -  \\
\hline \hline
\end{tabular}
\end{center}
\end{table*}

\begin{figure*}[htp]
{\psfig{file=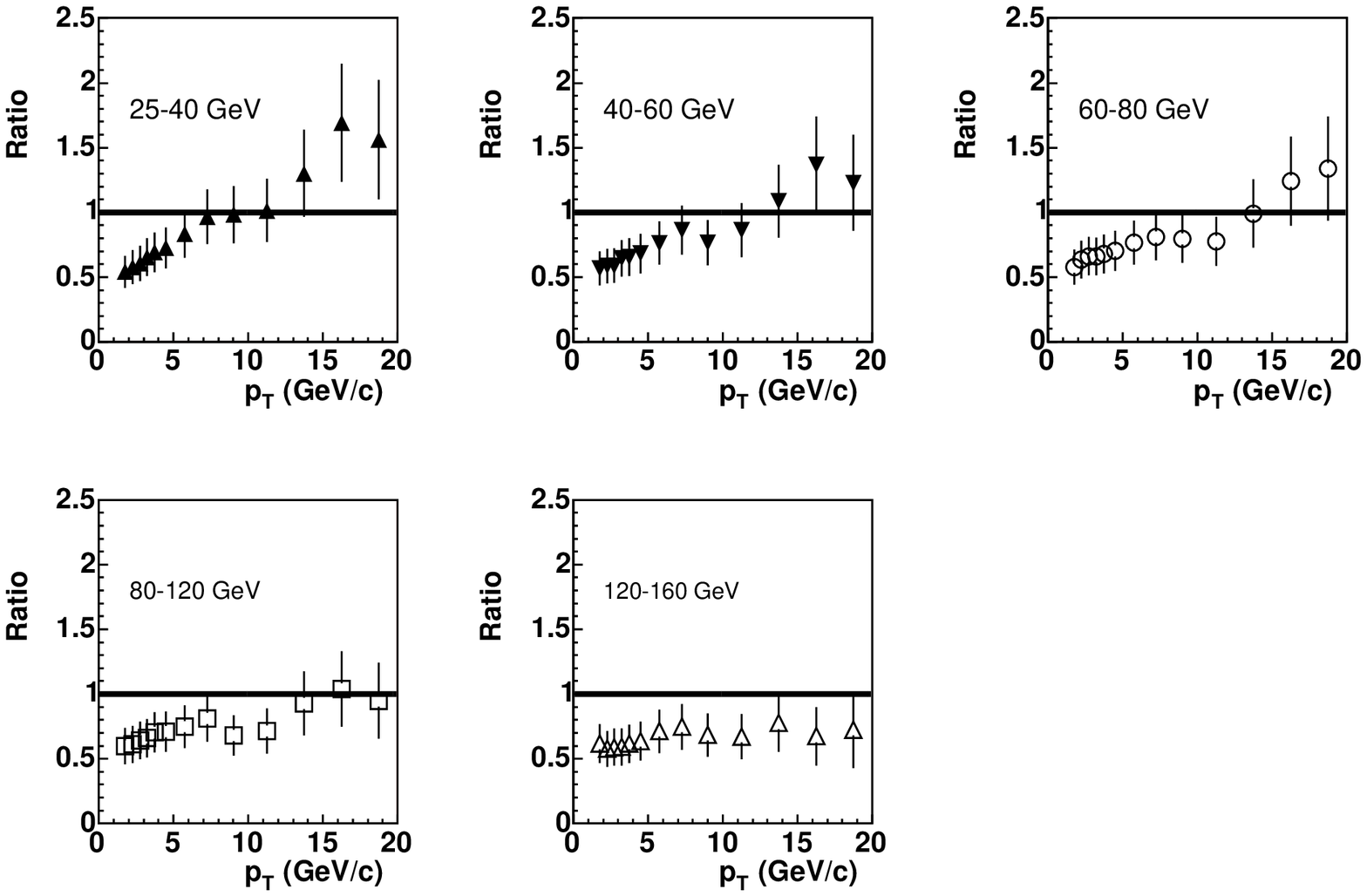,height=9.0cm}}
\caption{The ratios of \xkshort~$p_T$ distribution of data to that of {\sc pythia} (version 6) events generated
  with default parameters for the five jet-$E_T$ intervals. }
\label{fig:ratioksmc}
\end{figure*}

\begin{figure*}[htp]
{\psfig{file=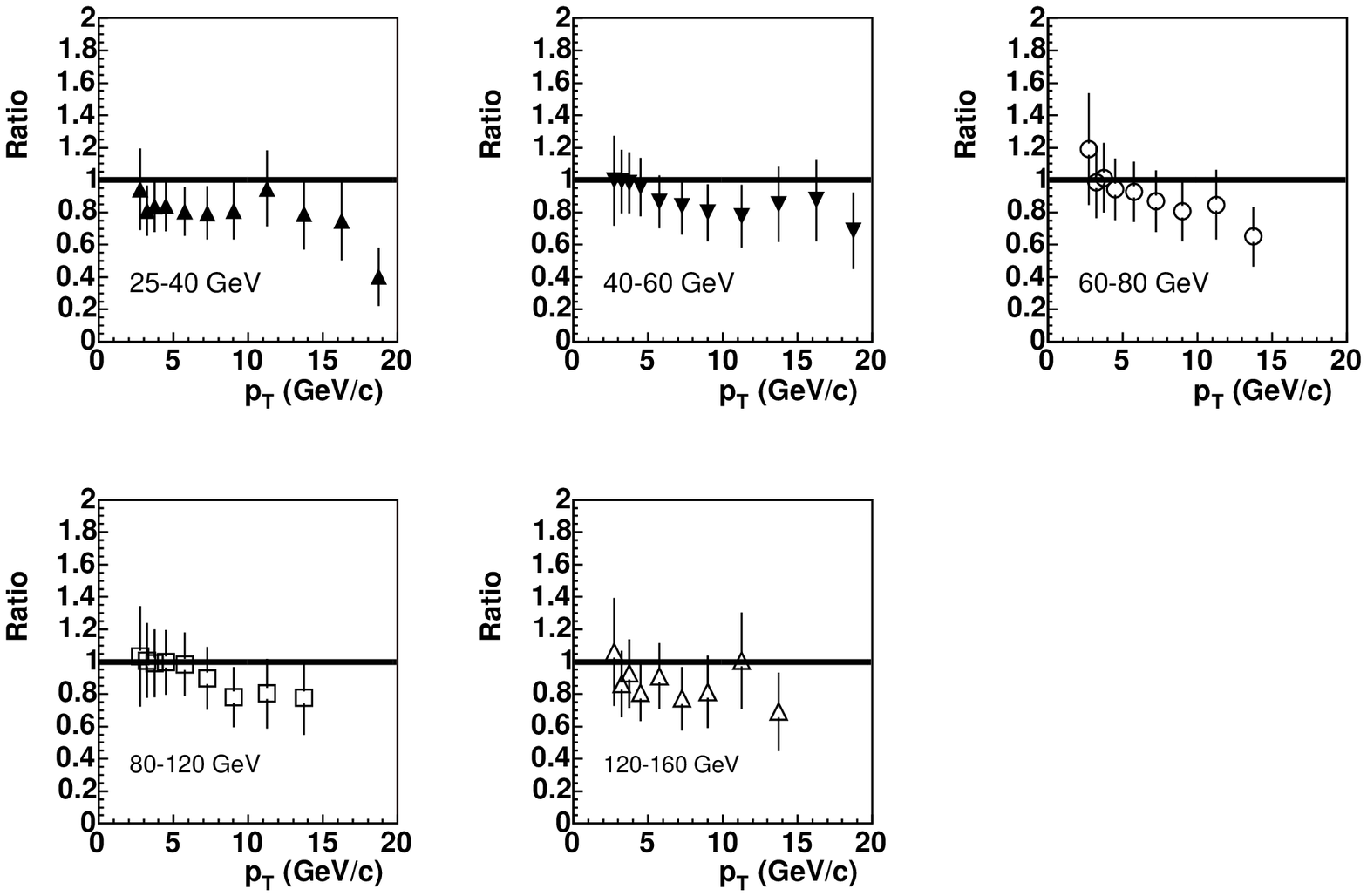,height=9.0cm}}
\caption{The ratios of \xlambda~$p_T$ distribution of data to that of {\sc pythia} (version 6) events generated  
with default parameters for the five jet-$E_T$ intervals. }
\label{fig:ratiolammc}
\end{figure*}

The differential $p_T$ distributions of \xkshort~and \xlambda~hadrons in jets are compared with the {\sc pythia} events 
generated with default parameters. The events from {\sc pythia} simulation, processed through CDF II 
detector simulation and reconstruction programs, are analyzed as the real data. However rather than finding the 
reconstructed number of resonances from fitting, the resonances at the particle generation level are 
used after associating them with reconstructed jets ($\Delta R<~$0.5) for the five jet-$E_T$ intervals. 
Figure~\ref{fig:ratioksmc} shows the ratios of the \xkshort~of data to that of {\sc pythia} 
events as a function of $p_T$, and Fig.~\ref{fig:ratiolammc} shows the same for \xlambda~baryons. The agreement for \xlambda~ 
baryons is adequate while {\sc pythia} generates too many \xkshort~mesons in the low $p_T$ region.

\begin{figure}[htp]
{\psfig{file=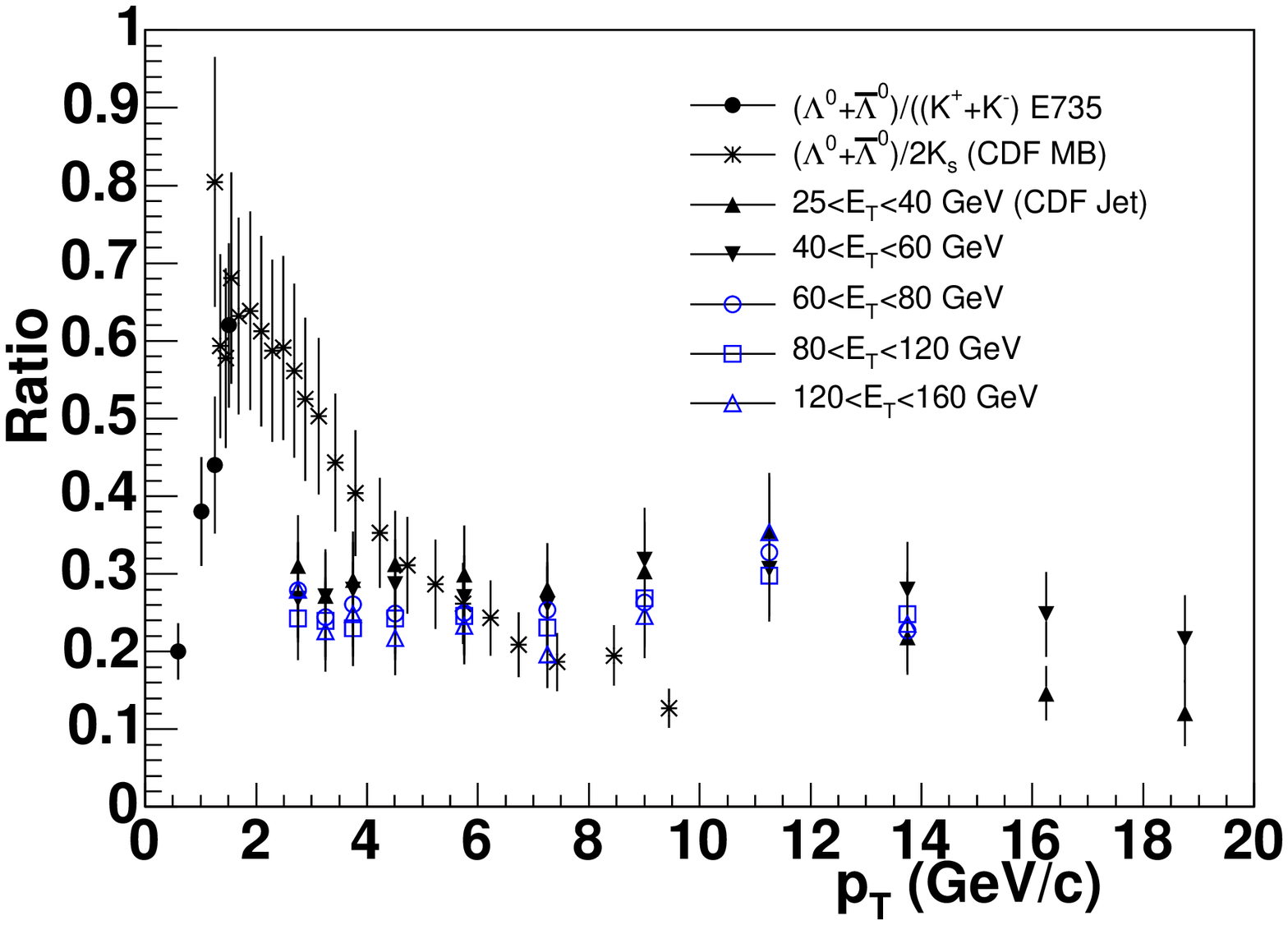,height=6.0cm}}
\caption{The cross-section ratios of \xlambda~baryon to strange meson as a function of $p_T$. The data from E735 is also from the 
   central region. The statistical and systematic uncertainties are added in quadrature.}
\label{fig:ratioall}
\end{figure}

Using the  $p_T$ distribution of \xlambda~baryons from an earlier 
analysis \cite{ref:cdfhyp}, and \xkshort~mesons from this analysis, the ratio of \xlambda~+ \xlambdab~to 
2\xkshort~as a function of $p_T$  in MB events is calculated and displayed in  Fig.~\ref{fig:ratioall}. In the same figure, 
the ratios from jets in Fig.~\ref{fig:ratiokl} are also shown for a comparison. Also shown in the figure 
is the ratio from 1.8 TeV center-of-mass energy covering the very low $p_T$ region ~\cite{ref:e735hyp}. 
The figure shows that the ratio of \xlambda~to \xkshort~exhibits different behavior than the  
$K^{*\pm}$ to $K_S^0$ and $\phi$ to $K_S^0$ ratios. For the latter, the ratios increase as $p_T$ 
increases and reache a plateau at $p_T>4 \sim~5$ GeV/$c$, while the former increases until $p_T$ reaches 
$\sim~2$ GeV/$c$ and then decreases as $p_T$ increases. The ratio plot also indicates that the process of 
producing \xlambda~baryons compared to \xkshort~mesons in MB events is significantly more efficient than the process in 
jets.  The ratio from the MB events matches the ratios from jets at $p_T$
$\sim~5$ GeV/$c$  implying that QCD jet contribution is significant for $p_T>5$ GeV/$c$.

\section{SUMMARY}

In inelastic $p \bar p$ collisions at $\sqrt{s} = 1.96$ TeV, we have studied the properties of three 
mesons, $K_S^0$, $K^{\star\pm}$, and $\phi$, in MB events  up to $p_T$ = 10 GeV/$c$ , and  
$K_S^0$ and $\Lambda^{0}$ hadrons in jets up to  20 GeV/$c$ $p_T$ and 160 GeV jet 
$E_T$. The measurements were made with centrally produced ($|\eta|<$~1) particles and jets. 
We found in MB events:\\*
1. As $p_T$ increases, the three mesons exhibit a similar $p_T$ slope 
as the $n$ values indicate, where $n$ is the exponent in the power law function. \\*
2. The $n$ values from the mesons are less than  the values from hyperons (\xlambda, $\Xi^\pm$, and $\Omega^\pm$) 
by about 10\%.  \\*
3. The ratios, \xkstar/2\xkshort~and \xphi/2\xkshort, as a function of $p_T$ exhibit a similar behavior as 
$\Xi^\pm$/(\xlambda + \xlambdab) and $\Omega^\pm$/(\xlambda +~\xlambdab)~\cite{ref:cdfhyp}. 
The ratios increase at low $p_T$ and reach a plateau above $p_T\sim$~5 GeV/$c$. \\*
4. Unlike the ratios among strange mesons or hyperons, the \xlambda~to \xkshort~ratio shows an enhancement 
around 2 GeV/$c$ $p_T$.  \\*
5. {\sc pythia} reproduces \xphi~$p_T$ distribution quite well, but underestimates \xkshort~production and overestimates
\xkstar~production. \\*
In jets:\\*
6. The $p_T$ dependencies of the \xkshort~and \xlambda~cross sections in jets reduce as jet $E_T$ increases. The ratio, 
(\xlambda + \xlambdab)/2\xkshort, is fairly constant at about 0.25 for $p_T$ up to 20 GeV/$c$ and jet $E_T$ 
up to 160 GeV. This ratio merges with the ratio from the MB events at $p_T > 4\sim$ 5 GeV/$c$. \\*
7. The process producing low $p_T$ \xlambda~(compared to \xkshort) in MB events is significantly more efficient 
than the process in jets. \\*
8. {\sc pythia} reproduces the \xlambda~$p_T$ distribution reasonably well, but overestimates \xkshort~production
in the low $p_T$ region. \\*
The findings indicate that in MB events particles with $p_T$ in excess of 5 GeV/$c$  are mostly from  QCD jets,
assuming that jets with $E_T <$ 25 GeV behave similarly to the higher $E_T$ jets.
The process of producing \xlambda~compared to \xkshort~around 2 GeV/$c$ is much more efficient 
than the process in jets. Moreover while the production cross section exhibits strong dependences on the 
quark flavors ($u$, $d$, and $s$) in particles, the $p_T$ slope  for $p_T>$ 5 GeV/$c$ is fairly insensitive 
to the number of quarks and quark flavors in particles, resulting in constant particle ratios. This  
suggests that $p_T$ dependences of particles produced in jets are similar regardless of their quark and flavor 
content. 

\clearpage
\newpage

We thank the Fermilab staff and the technical staffs of the participating institutions for their vital contributions. This work was supported by the U.S. Department of Energy and National Science Foundation; the Italian Istituto Nazionale di Fisica Nucleare; the Ministry of Education, Culture, Sports, Science and Technology of Japan; the Natural Sciences and Engineering Research Council of Canada; the National Science Council of the Republic of China; the Swiss National Science Foundation; the A.P. Sloan Foundation; the Bundesministerium f\"ur Bildung und Forschung, Germany; the Korean World Class University Program, the National Research Foundation of Korea; the Science and Technology Facilities Council and the Royal Society, UK; the Russian Foundation for Basic Research; the Ministerio de Ciencia e Innovaci\'{o}n, and Programa Consolider-Ingenio 2010, Spain; the Slovak R\&D Agency; the Academy of Finland; and the Australian Research Council (ARC).



\begin{thebibliography}{99}

\bibitem{ref:isrjet}
T.~Akesson {\it et al.} (AFS Collaboration) Phys. Lett. B {\bf 118}, 185 (1982).

\bibitem{ref:ua1jet}
M.~Banner {\it et al.} (UA2 Collaboration) Phys. Lett. B {\bf 118}, 203 (1982);
G.~Arnison {\it et al.} (UA1 collaboration), Phys. Lett. B {\bf 123}, 115 (1983).

\bibitem{ref:hhcollision}
 G.J.~Alner {\it et al.} (UA5 Collaboration), Nucl. Phys. {\bf B258}, 505 (1985);
 R.E.~Ansorge {\it et al.} (UA5 Collaboration), Phys. Lett. B {\bf 199}, 311 (1987);
 T.~Alexopolos {\it et al.} (E735 Collaboration) , Phys. Rev. Lett. {\bf 60}, 1622 (1988);
 F.~Abe {\it et al.}  (CDF Collaboration), Phys. Rev. Lett. {\bf 61}, 1819 (1988);
 R.~Ansoge {\it et al.} (UA5 Collaboration), Nucl. Phys.  {\bf B328}, 36 (1989);
 G.~Bocquet {\it et al.} (UA1 Collaboration), Phys. Lett.  B {\bf 366}, 441 (1996);
 S.~Banerjee {\it et al.} (E735 Collaboration), Phys. Rev. Lett.  {\bf 64}, 991 (1990);
 D.~Acosta {\it et al.} (CDF Collaboration), Phys. Rev. D {\bf 72}, 052001 (2005);
 B.~I.~Abelev {\it et al.} (STAR Collaboration), Phys. Rev. C {\bf 75}, 064901, (2007).

\bibitem{ref:pythia} 
  T.~Sj\"{o}strand, P.~Eden, C.~Friberg, L.~Lonnblad, G.~Miu, S.~Mrenna, and E.~Norrbin,
  Comput.\ Phys.\ Commun.\  {\bf 135}, 238 (2001). The version used in this paper is 6.327.
 
\bibitem{ref:eedata}
 W.~Braunschweig {\it et al.} (TASSO Collaboration), Z.\ Phys. C {\bf 47}, 167 (1990);
 H.~Aihara {\it et al.} (TPC Collaboration), Phys.\ Rev.\ Lett.\ {\bf 54}, 274 (1985);
 H.~Schellman {\it et al.} (MARK-II Collaboration), Phys.\ Rev. D {\bf 31}, 3013 (1985);
 M.~Derrick {\it et al.} (HRS Collaboration), Phys.\ Rev.D {\bf 35}, 2639 (1987);
 H.~Behrend {\it et al.} (CELLO Collaboration), Z.\ Phys. C {\bf 46}, 397 (1990);
 D.~Buskulic {\it et al.} (ALEPH Collaboration),  Z.\ Phys. C {\bf 64}, 361 (1994);
 P.~Abreu {\it et al.} (DELPHI Collaboration), Z.\ Phys. C {\bf 65}, 587 (1995);
 M.~Acciarri {\it et al.} (L3 Collaboration), Phys.\ Lett. B {\bf 328}, 223 (1994);
 P.~Acton {\it et al.} (OPAL Collaboration), Phys.\ Lett. B {\bf 291}, 503 (1992).
 
\bibitem{ref:epdata}
 M.~Derrick {\it et al.} (ZEUS Collaboration), Z.\ Phys. C {\bf 68}, 29 (1995);
 S.~Chekanov {\it et al.} (ZEUS Collaboration), Eur.\ Phys.\ J. C {\bf  51}, 1, (2001);
 S.~Aid {\it et al.} (H1 Collaboration), Nucl.\ Phys.\ {\bf B480}, 3 (1996);
 C.~Adloff {\it et al.} (H1 Collaboration), Z.\ Phys. C {\bf 76}, 213 (1997).

\bibitem{ref:cdfdet} 
  A.~Abulencia {\it et al.}  (CDF Collaboration), J.\ Phys. G {\bf 34}, 2457 (2007).

\bibitem{ref:cdfsvx}
A.~Sill {\it et al.}, Nucl.\ Instrum.\ Methods\ A {\bf 447}, 1 (2000).

\bibitem{ref:cdfxyz} 
 In the CDF coordinate system, $\theta$ and $\phi$ are the polar and azimuthal
 angles of a particle, respectively, defined with respect to the proton beam
 direction, $z$. The pseudorapidity $\eta$ is defined as $-\ln [\tan(\theta/2)]$.
 The transverse-momentum of a particle is $p_T = p\sin{\theta}$. The rapidity 
 is defined as $y = 0.5 \ln [ (E+p_{z})/(E-p_{z}) ]$, where $E$ and $p_z$ are the 
 energy and longitudinal momentum of the particle.

\bibitem{ref:cdfcot}
A.~Af\mbox{}folder {\it et al.}~(CDF Collaboration), Nucl.\ Instrum.\ Methods\ A {\bf 526}, 249 (2004).


\bibitem{ref:cdfcal1}
L. Balka {\it et al.}, Nucl. Instrum. Methods  A {\bf 267}, 272 (1988); 
S. Bertolucci {\it et al.}, Nucl. Instrum. Methods  A {\bf 267}, 301 (1988); 
M. Albrow {\it et al.}, Nucl. Instrum. Methods  A {\bf 480}, 524 (2002).

\bibitem{ref:cdfcal2}
G. Apollinari {\it et al.}, Nucl. Instrum. Methods A {\bf 412}, 515 (1998).

\bibitem{ref:cdfclc}
D.~Acosta {\it et al.}, Nucl.\ Instrum.\ Methods\ A {\bf 494}, 57 (2002).

\bibitem{ref:cdfcal3}
A.~Bhatti {\it et al.}, Nucl. Instrum. Methods A {\bf 566}, 375 (2006).

\bibitem{ref:cdfhyp} 
T.~Aaltonen {\it et al.}~(CDF Collaboration), Phy. Rev. D {\bf 86}, 012002 (2012).
The $n$ values are 8.81 $\pm$ 0.08 (\xlambda + \xlambdab), 8.26 $\pm$ 0.12 ($\Xi^\pm$) 
and 8.06 $\pm$ 0.34 ($\Omega^\pm$).


\bibitem{GEANT}
  R.~Brun, R.~Hagelberg, M.~Hansroul, and J.~C.~Lassalle,
  version 3.15, CERN-DD-78-2-REV.

\bibitem{ref:pdg} 
  C.~Amsler {\it et al.}  (Particle Data Group),
  Phys.\ Lett. B {\bf 667}, 1 (2008).

\bibitem{ref:ndxsection} 
The total cross section corresponding to the MB trigger is estimated 
to be $45\pm 8$ mb. The elastic ($17\pm4$ mb~\cite{ref:pdg}), single diffractive SD 
($12$ mb), and half of the double diffractive DD ($4$ mb) cross sections are 
subtracted from the total $p\overline{p}$ cross section ($78\pm6$ mb~\cite{ref:pdg}~\cite{ref:compete}) 
to give this estimate.  The SD and DD cross sections are estimated using {\sc pythia}~\cite{ref:pythia},
and no uncertainties are assigned to SD and DD cross section.
A simulation study shows that the MB trigger is sensitive to $\sim~100$\% 
of inelastic events which are not SD or DD and $\sim~50$\% of DD events.  
A 100\% uncertainty is assigned to the DD contribution due to the uncertainty 
in the event characteristics and detector simulation.

\bibitem{ref:compete} 
  J.R. Cudell {\it et al.}  (COMPETE Collaboration), Phys. Rev. Lett. {\bf 89}, 201801 (2002)

\bibitem{ref:jpsi} 
  D. Acosta {\it et al.}~(CDF Collaboration), Phys. Rev. D {\bf 71}, 032001 (2005).

\bibitem{ref:upsilon} 
  F. Abe {\it et al.}~(CDF Collaboration), Phys. Rev. Lett. {\bf 75}, 4358 (1995).

\bibitem{ref:cdfks} 
  F.~Abe {\it et al.}~(CDF Collaboration),
  Phys.\ Rev. D {\bf 40}, 3791 (1989).
  
\bibitem{ref:private} 
 Private communication with one of {\sc pythia}~authors.

\bibitem{ref:e735phi}
  T.~Alexopoulos {\it et al.}  (E735 Collaboration),
   Z. Phys. C {\bf 67}, 411416 (1995).

\bibitem{ref:e735hyp} 
  S.~Banerjee {\it et al.}  (E735 Collaboration),
  Phys.\ Rev.\ Lett.\  {\bf 62}, 12 (1989).
  T.~Alexopoulos {\it et al.}  (E735 Collaboration),
  Phys.\ Rev. D {\bf 46}, 2773 (1992).



\end{thebibliography}
\end{document}